\documentclass[9pt,twocolumn,twoside]{optica}
\setboolean{shortarticle}{false}
\setboolean{minireview}{false}

\usepackage{mathtools}
\usepackage{siunitx}
\usepackage{stmaryrd}
\title{Common pulse retrieval algorithm: a fast and universal method to retrieve ultrashort pulses}

\author[1,*]{Nils C. Geib}
\author[1]{Matthias Zilk}
\author[1,2,3]{Thomas Pertsch}
\author[1,2,3]{Falk Eilenberger}

\affil[1]{Institute of Applied Physics, Abbe Center of Photonics, Friedrich Schiller University, Albert-Einstein-Str. 15, 07745 Jena, Germany}
\affil[2]{Fraunhofer Institute for Applied Optics and Precision Engineering IOF, Center for Excellence in Photonics, Albert-Einstein-Str. 7, 07745 Jena, Germany}
\affil[3]{Max Planck School of Photonics, Germany}

\affil[*]{Corresponding author: nils.geib@uni-jena.de}

\makeatletter
\renewcommand{\p@subsection}{\thesection.}
\makeatother

\begin{abstract}
We present a common pulse retrieval algorithm~(COPRA) that can be used for a broad category of ultrashort laser pulse measurement schemes including frequency-resolved optical gating~(FROG), interferometric FROG, dispersion scan, time domain ptychography, and pulse shaper assisted techniques such as multiphoton intrapulse interference phase scan~(MIIPS). We demonstrate its properties in comprehensive numerical tests and show that it is fast, reliable and accurate in the presence of Gaussian noise. For FROG it outperforms retrieval algorithms based on generalized projections and ptychography. Furthermore, we discuss the pulse retrieval problem as a nonlinear least-squares problem and demonstrate the importance of obtaining a least-squares solution for noisy data. These results improve and extend the possibilities of numerical pulse retrieval. COPRA is faster and provides more accurate results in comparison to existing retrieval algorithms. Furthermore, it enables full pulse retrieval from measurements for which no retrieval algorithm was known before, e.g.,~MIIPS measurements. 
\end{abstract}

\setboolean{displaycopyright}{true}

\newcommand{\wb}{\omega}
\newcommand{\w}{\varOmega}
\newcommand{\env}{E}
\newcommand{\tenv}{\env}
\newcommand{\senv}{\tilde{\env}}
\newcommand{\Hf}{\tilde{H}}
\newcommand{\envt}{\tenv(t)}
\newcommand{\envw}{\senv(\wb)}
\newcommand{\inv}{^{\raisebox{.2ex}{$\scriptscriptstyle-1$}}}
\newcommand{\F}{\mathcal{F}\mathopen{}}
\newcommand{\Fi}{\mathcal{F}\inv\mathopen{}}
\newcommand{\intiti}{\int_{-\infty}^{\infty}}
\newcommand{\D}{\mathrm{d}}
\newcommand{\e}{\mathrm{e}}
\newcommand{\im}{\mathrm{i}}
\newcommand{\Si}{\mathcal{S}\mathopen{}}
\newcommand{\N}{\mathcal{N}\mathopen{}}
\newcommand{\vE}{\mathbf{\senv}}
\newcommand{\Tm}{T^{\mathrm{meas}}}
\newcommand{\dft}[2]{\mathrm{FT}_{{#2}\shortrightarrow{#1}\mathopen{}}}
\newcommand{\idft}[2]{\mathrm{FT}_{{#2}\shortrightarrow{#1}\mathopen{}}\inv}
\newcommand{\nnR}{r}

\renewcommand{\epsilon}{\varepsilon}

\begin{document}

\maketitle
\thispagestyle{fancy}
\ifthenelse{\boolean{shortarticle}}{\abscontent}{}

\section{Introduction}\label{sec:introduction}
Since the advent of ultrashort laser pulses there has been ongoing research on techniques to determine their temporal structure. 
Nowadays, there is quite literally a "zoo" of techniques available for that purpose~\cite{Walmsley2009}.

The direct measurement of the temporal intensity of laser pulses using electrical detectors is limited to the picosecond range due to their relatively slow response time. Autocorrelation measurements~\cite{Ippen1977, Diels1985} were introduced to overcome this limitation and are still the most widely used pulse characterization methods.
However, it is not possible to retrieve the full pulse information from a single autocorrelation measurement, as it is ambiguous with respect to the pulse amplitude and phase~\cite{Chung2001}.

A prominent method, which enables the reconstruction of both the pulse amplitude and phase, is called frequency-resolved optical gating~(FROG)~\cite{Kane1993,Trebino2002}. 
It extends the non-collinear intensity autocorrelation by measuring the spectrum of its nonlinear signal for every delay. 
The most common variant of FROG utilizes non-collinear second harmonic generation~(SHG). 
The resulting two-dimensional measurement, a set of frequency-doubled spectra, is called the SHG-FROG trace. It is presumed to uniquely define both pulse amplitude and phase except for certain, so-called trivial, ambiguities~\cite{Seifert2004,Bendory2017}.

Several variants of FROG exist that use other nonlinear processes such as third-harmonic generation~(THG), self-diffraction~(SD) and polarization gating~(PG)~\cite{Trebino2002}. 
An interferometric variant of SHG-FROG that is based on the collinear autocorrelation is called interferometric FROG (iFROG)~\cite{Stibenz2005}. Recently it has been demonstrated by using THG as the nonlinear process~\cite{Hyyti2017a,Hyyti2017b}.

Reconstructing a pulse from a FROG trace measurement requires an iterative algorithm. 
One successful approach was inspired by projection on convex sets~\cite{Bauschke1996} and is called the generalized projections algorithm~(GPA)~\cite{DeLong1994}. 
An improved version that exploits the specific algebraic structure of a FROG trace for faster retrieval is called the principal components generalized projections algorithm~(PCGPA)~\cite{Kane1997}.

A more recent pulse measurement technique is dispersion scan~(d-scan) which has become a valuable tool for few-cycle pulse measurement~\cite{Miranda2012,Miranda2012a}.
In this method the pulses are chirped by inserting an adjustable dispersive element, e.g., a pair of glass wedges, into the beam path and their SHG spectrum is measured as a function of the induced chirp to form the d-scan trace. 
The pulse can then be retrieved from the trace by using a multi-dimensional optimization algorithm. 
The technique was also demonstrated using THG and inline SD as the nonlinear process~\cite{Hoffmann2014,Canhota2017}. 
Recently a fast, iterative algorithm based on generalized projections was proposed to enable pulse retrieval from SHG and THG d-scan traces~\cite{Miranda2017}.

A third class of pulse measurement methods can be implemented using a pulse shaper.
With this a set of spectral phase masks is applied to the pulses.
Then the SHG spectrum for every phase mask is measured to obtain a two-dimensional measurement trace.
The most prominent technique in this class is called multiphoton intrapulse interference phase scan (MIIPS)~\cite{Lozovoy2004,Xu2006}, where sinusoidal phase patterns with varying shifts are applied.
So far, only iterative spectral phase compensation has been demonstrated by using an algorithm that extracts an approximation to the second derivative of the spectral phase from the measurement.
Other techniques represent adaptations of existing measurement schemes to the use with a pulse shaper~\cite{Galler2008,Forget2010,Loriot2013,Wilcox2014}.

Time-domain ptychography~(TDP) is a recently developed pulse measurement technique~\cite{Spangenberg2015,Spangenberg2016,Witting2016}. It is inspired by a coherent diffractive imaging technique of the same name~\cite{Hegerl1970}.
It uses a correlation setup similar to FROG where the pulse in one arm is spectrally filtered. 
The pulse retrieval algorithms used for TDP are an adaption of the image retrieval algorithms in spatial ptychography~\cite{Maiden2009,Maiden2017}. 
They have also been successfully applied to cross-correlation FROG (XFROG)~\cite{Heidt2016} and SHG-FROG~\cite{Sidorenko2016,Sidorenko2017}.

Typically, each family of these pulse measurement methods comprises both a specific experimental setup and a taylored retrieval algorithm, which makes them difficult to compare. They are, however, structurally similar. For that reason, our paper starts in Sec.~\ref{sec:concepts} by developing a common formalism for what we call parametrized nonlinear process spectra~(PNPS) measurements. It allows to describe most self-referenced techniques for ultrashort pulse measurement using the same formalism. It forms the basis of our work and is used to develop all arguments in the following sections.

The main idea of our paper is presented in Sec.~\ref{sec:pulse_retrieval_problem}. We discuss PNPS pulse retrieval as a nonlinear least-squares problem. We propose this as the natural way to view the pulse retrieval problem and stress that the least-squares solution is ideal under the assumption of Gaussian noise.

The main result of our paper is the common pulse retrieval algorithm~(COPRA) described in Sec.~\ref{sec:algorithm}. It can be applied universally to all PNPS measurements. It is in general faster than general least-squares solvers and more accurate than other specialized pulse retrieval algorithms. It significantly extends the possibilities of numerical pulse retrieval because for some PNPS measurements no fast retrieval algorithm was known before (e.g., SD-dscan) and for some no phase and amplitude retrieval algorithm existed at all (e.g., MIIPS). 

In Secs.~\ref{sec:methods} and \ref{sec:results} we describe the comprehensive numerical tests that were performed to verify and demonstrate the properties of COPRA. They included testing the retrieval for various PNPS measurements, test pulses and levels of noise. For SHG-FROG we compare it to PCGPA and ptychographic retrieval. We can show that those algorithms do not converge on a least-squares solution and, consequently, are much less accurate in the presence of Gaussian measurement noise. We also demonstrate that COPRA is able to retrieve pulses from incomplete traces. Furthermore, we evaluate the applicability of general minimization algorithms on the pulse retrieval problem. We find that gradient-based algorithms such as Levenberg-Marquadt are generally superior to gradient-free methods.

In Sec.~\ref{sec:conclusion} we summarize the results and give an outlook on future work. In the supplementary material we also give more details and technical aspects that facilitate the application and reimplementation of COPRA.

\section{Concepts}\label{sec:concepts}
In this section we introduce a unified description of most self-referenced pulse measurement methods. It is based on the observation that the measured quantity is the same for all methods mentioned in the introduction: a set of pulse spectra after a nonlinear process.
Specifically, the nonlinear process is tunable by some parameter that forms the second measurement dimension. 
For example, for SHG-FROG the parameter is the pulse delay and the nonlinear process a non-collinear SHG. For SHG-d-scan the parameter is the insertion distance of a glass wedge and the nonlinear process is a collinear SHG. 

We call these measurements parametrized nonlinear process spectra~(PNPS) measurements. Other pulse measurement techniques exist that cannot be described in this way. 
Most prominently this pertains to spectral phase-interferometry for direct electric field reconstruction (SPIDER)~\cite{Iaconis1998} and other methods based on spectral interferometry. 
They do not require a retrieval algorithm and are not subject of this paper.

\subsection{Continuous PNPS formalism}
We work with the complex-valued pulse envelope $\envt$ and its spectral counterpart $\envw$, where $\wb = \w - \w_0$ is the centered frequency and $\w_0$ the central frequency.
Both are related by the Fourier transform and its inverse using the following convention
\begin{align}
\envw &= \F\left[ \tenv \right](\wb) = \frac{1}{2\pi} \intiti \envt \:\e^{\im \wb t} \D t, \\
\envt &= \Fi\left[\senv \right](t) = \intiti \envw \: \e^{-\im t \wb} \D \wb.
\end{align}
All PNPS \emph{traces} $T$ can be modeled by the following equation
\begin{align}
 T(\delta, \wb; \senv) = \left|\F\left\{\Si_\delta \left[\senv\right](t) \right\}(\wb)\right|^2. \label{eq:pnps_1}
\end{align}
$T$ depends on the pulse $\senv$ and is evaluated at the frequency $\wb$ and the parameter $\delta$. $\Si_\delta$ is the \emph{signal operator} that describes a parametrized nonlinear process in the time domain. $\delta$ is a method-specific parameter that tunes the nonlinear process.

Depending on the structure of the signal operator we distinguish between non-collinear methods~(e.g., FROG or TDP) and collinear methods~(e.g., d-scan or iFROG). Examples of the signal operator in the former case can be found in Tab.~\ref{tab:measurement_operators}. In the latter case we can decompose the signal operator in the following way
\begin{align}
\Si_\delta[\senv](t) = \N\left\{\Fi\left[\Hf_\delta \: \senv \right]\right\}(t). \label{eq:pnps_2}
\end{align}
$\Hf_\delta(\wb)$ is the \emph{parametrization filter} and describes a parametrized linear operation in the frequency domain. Examples are listed in Tab.~\ref{tab:parametrization_filters}.
$\N$ is the \emph{nonlinear process operator} and describes the subsequent conversion of a pulse by a collinear nonlinear process. Expressions for the processes commonly used in pulse measurement can be found in Tab.~\ref{tab:nonlinear_operators}.

PNPS traces do not uniquely define a pulse. For example, they are all ambiguous to the constant and linear phase of $\senv(\wb)$. Some methods (e.g., SHG-FROG and SHG-iFROG) leave the direction of time undetermined.
Additionally, the relative phase of pulse components well-separated in frequency can be shown to be ambiguous, similar to how it was done for FROG~\cite{Keusters2003a}.
Answering the underlying question if a PNPS trace is essentially unique, i.e., if it defines pulse amplitude and phase up to a set of known, so-called trivial ambiguities, is out of scope for this work.
Even for the well-studied FROG method it is still a topic of ongoing research~\cite{Seifert2004,Bendory2017,Bendory2017a}.
We will take a pragmatic approach and test for non-trivial ambiguities by numerically retrieving pulses from a large number of synthetic measurements.

\begin{table}[t]
\begin{center}
\caption{\bf Signal operator for selected non-collinear schemes.}
\begin{tabular}{cl}
\hline
Method & $\Si_\tau[\senv]$ \\
\hline
SHG-FROG &  $\F^{-1}\left[\e^{\im \tau \wb} \senv \right] \: \F^{-1}\left[\senv \right]$ \\
PG-FROG & $\left|\F^{-1} \left[\e^{\im \tau \wb} \senv \right] \right|^2 \: \F^{-1}\left[\senv \right]$ \\
TDP\textsuperscript{a} & $\F^{-1} \left[\e^{\im \tau \wb} \tilde{B}(\wb) \senv \right]  \: \F^{-1}\left[\senv \right]$ \\
\hline
\end{tabular}
\label{tab:measurement_operators}
\end{center}
{\small The pulse delay $\tau$ is the parameter $\delta$ in these methods. More examples can be found in the supplementary material. \textsuperscript{a}$\tilde{B}(\wb)$ describes the transmission of a bandpass filter used in the scheme.}
\end{table}
\begin{table}[t]
\begin{center}
\caption{\bf Parametrization filter for selected collinear schemes.}
\begin{tabular}{lp{2.5cm}l}
\hline
Scheme & Parameter $\delta$ & $\Hf_\delta(\wb)$  \\
\hline
d-scan\textsuperscript{a} & glass insertion $z$ & $\exp[\im k(\wb + \w_0) z]$ \\
MIIPS\textsuperscript{b} & pattern shift $\delta$ & $\exp[\im \alpha \cos(\gamma \wb - \delta)]$ \\    
iFROG& delay $\tau$ & $1/2 + \exp[\im \tau (\wb + \w_0)]/2$ \\
\hline
\end{tabular}\label{tab:parametrization_filters}
\end{center}
{\small More examples can be found in the supplementary material. \noindent{}\textsuperscript{a}$k(\w)$ depends on the material of the wedges and is usually defined by Sellmeier equations. \textsuperscript{b}$\alpha$ and $\gamma$ are free parameters of the method and have to be adapted to the measured pulses.}
\end{table}
\begin{table}[t]
\begin{center}
\caption{\bf Nonlinear process operators for collinear schemes.}
\begin{tabular}{cccc}
\hline
Process & SHG & THG & SD \\
\hline
$\N[\tenv]$ & $\tenv^2$ & $\tenv^3$ & $\left|\tenv\right|^2 \tenv$ \\
\hline
\end{tabular}
\label{tab:nonlinear_operators}
\end{center}
\end{table}
\subsection{Discrete PNPS formalism}\label{subsec:discrete_formalism}
To perform pulse retrieval we have to a introduce a discrete version of the PNPS formalism. We define equidistant simulation grids with $N$ points in time and frequency
\begin{align}
t_n &\equiv t_0 + n \: \Delta t, \quad n=0, \ldots, N-1 \\
\wb_n &\equiv \wb_0 + n \: \Delta \wb.
\end{align}
We set $\tenv_n \equiv \tenv(t_n)$ and $\senv_n \equiv \senv(\wb_n)$. We use $\mathbf{E} \equiv (\tenv_0, \ldots, \tenv_{N-1})$ and $\vE \equiv (\senv_0, \ldots, \senv_{N-1})$ to denote the whole pulse. The Fourier transform is approximated by discrete evaluation of the integral and is denoted by
\begin{align}
 \senv_n = \dft{n}{k}(\tenv_k) \quad \text{and} \quad 
 \tenv_k = \idft{k}{n}(\senv_n).
\end{align}
We have $M$ spectra for the parameters $\delta_0, \ldots, \delta_{M-1}$. There is no restriction on the number, spacing or position of the tuning parameters $\delta_m$, e.g., as it is required by PCGPA for FROG. The discrete PNPS signal $S_{mk}$ is defined by a discrete evaluation of the signal operator at $\delta_m$ and $t_k$:
\begin{align}
 S_{mk} \equiv S_{mk}(\vE) \approx \Si_{\delta_m}[\senv](t_k). \quad \begin{split} m&=0, \ldots, M-1 \\ k&=0, \ldots, N-1 \end{split} \label{eq:complex_measurement_trace}
\end{align}
Its counterpart in the frequency domain is denoted by
\begin{align}
 \tilde{S}_{mn} \equiv  \tilde{S}_{mn}(\vE) = \dft{n}{k}(S_{mk}), \quad \begin{split} m&=0, \ldots, M-1\\ n&=0, \ldots, N-1\end{split}\label{eq:complex_measurement_spectrum}
\end{align}
Finally, we can calculate the discrete PNPS trace $T_{mn}$ by
\begin{align}
    T_{mn} \equiv  T_{mn}(\vE) = |\tilde{S}_{mn}(\vE)|^2 \approx T(\delta_m, \wb_n; \senv). \label{eq:measurement_trace}
\end{align}
The measurement from which the pulse is reconstructed, the measured PNPS trace, is denoted by $\Tm_{mn}$. More details on how the calculations are performed can be found in the supplementary material~(Sec.~S2).

\section{Pulse retrieval problem}\label{sec:pulse_retrieval_problem}
\begin{figure*}[tb]
	\centering
	\includegraphics{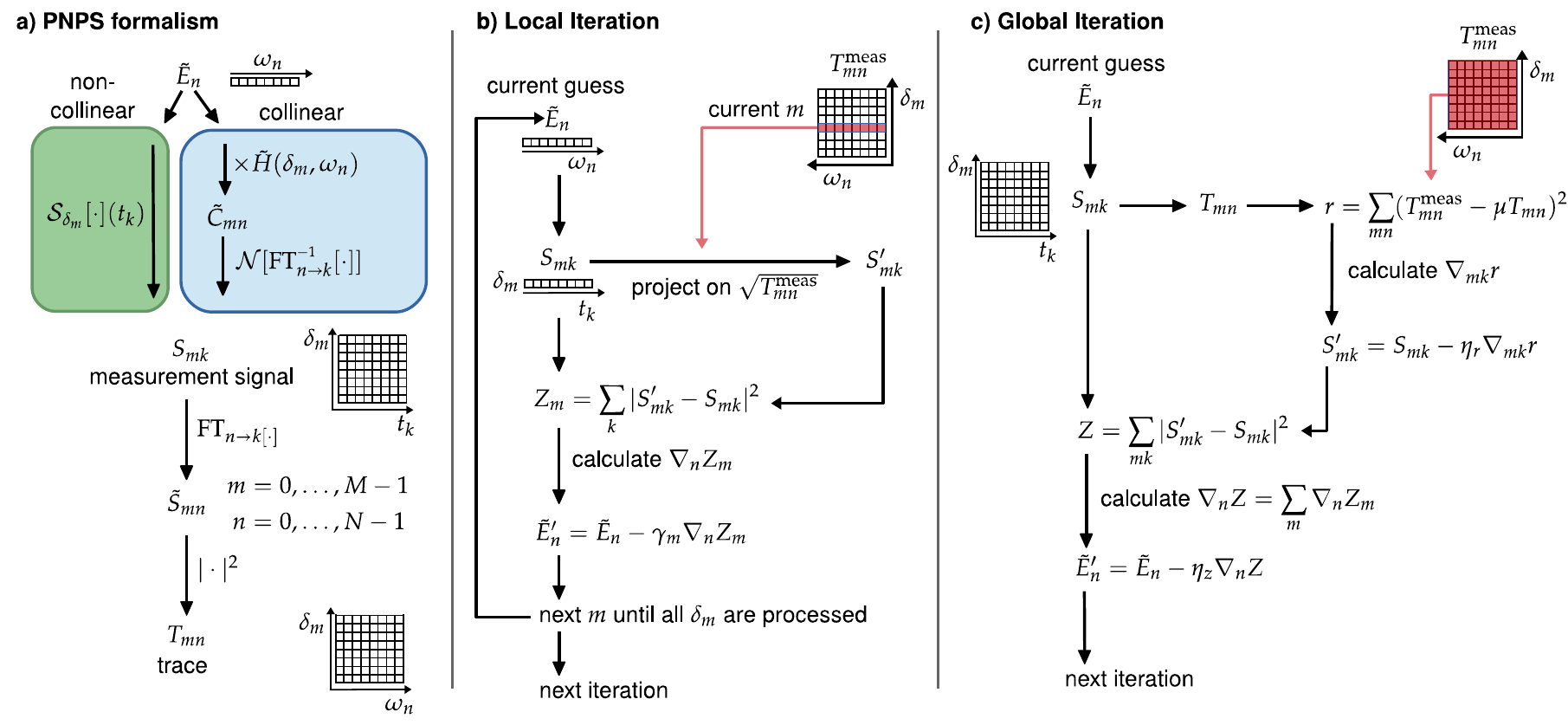}
	\caption{a) A diagram of the discrete PNPS formalism. b) First stage of COPRA: local iteration c) Second stage of COPRA: global iteration.}
	\label{fig:algorithm}
\end{figure*}
The discrete pulse retrieval problem is to find the pulse $\vE$ that gives rise to a PNPS trace $T_{mn}$ that matches the measurement $\Tm_{mn}$. As $\Tm_{mn}$ is subject to measurement errors the retrieval will never be exact and we need to choose a metric to select the best solution. 

In this work we view pulse retrieval as a nonlinear least-squares problem. It is a nonlinear inverse problem and solving such problems in the least-squares sense is very well-established and understood~\cite{Tarantola2005}. Under the assumption of Gaussian measurement errors the least-squares solution represents an optimal choice, namely the maximum-likelihood estimate~\cite{Seber2003} and a Gaussian distribution is usually a good model for noise in spectrometric measurements~\cite{Davenport2015}.

To explain why we put so much emphasis on finding a least-squares solution we have to anticipate a key result from Sec.~\ref{subsec:copra}. We found that retrieval algorithms based on generalized projections or ptychography do not converge onto a solution in the least-squares sense. These methods use the sum of squared residuals, e.g., the FROG error $G$, to assess and select a solution. However, they minimize it only approximately. In consequence, they under-perform in the presence of additive Gaussian noise. This fact has been noticed several times in the literature~\cite{Wilcox2014,Miranda2017,Hyyti2017b,Escoto2018}, but the relation to the missing least-squares property was not reported so far.

With this in mind we state the pulse retrieval problem as fitting the $2N$ independent variables in $\vE$ (real and imaginary parts) to $MN$ dependent variables in $\Tm_{mn}$ by minimizing the sum of squared residuals $r$
\begin{align}
 r \equiv r(\vE) = \sum_{m,n} \left[\Tm_{mn} - \mu T_{mn}(\vE) \right]^2. \label{eq:residuals}
\end{align}
Throughout the paper we use the \emph{trace error} $R$ defined by
\begin{align}
 R \equiv R(\vE) = r^{1/2} / \bigl[ MN \: ( \operatorname*{max}_{m,n} \Tm_{mn})^2 \bigr]^{1/2} \label{eq:data_error}
\end{align}
to assess the convergence. $R$ is the normalized root-mean-square error~(NRMSE) between $\Tm_{mn}$ and $\mu T_{mn}$. The normalization facilitates the comparison of $R$ between different measurements and PNPS schemes. In the FROG literature $\max \Tm = 1$ is usually assumed implicitly and, thus, $R$ is then equivalent to the FROG error $G$.

The scaling factor $\mu$ in \eqref{eq:residuals} accounts for different scales of the measured and computed traces. Its value can be obtained for every $\vE$ from an analytical solution by
\begin{align}
\mu = \sum_{m,n} \left[ \Tm_{mn} \: T_{mn}(\vE) \right] / \sum_{m,n} T_{mn}(\vE)^2 . \label{eq:scaling_factor}
\end{align}
A general solution strategy for the pulse retrieval problem is to minimize $r$ by employing a nonlinear minimization algorithm as demonstrated in Sec.~\ref{subsec:nonlinear_solvers}. However, in general such an approach will be less efficient than a specialized algorithm like the one we present in the next section.

\section{Common pulse retrieval algorithm}\label{sec:algorithm}
In the following we present a fast iterative pulse retrieval algorithm that is able to solve the PNPS pulse retrieval problem in the least-squares sense. We call it the common pulse retrieval algorithm~(COPRA).

Once the algorithm is implemented it can easily be applied to a multitude of present and future PNPS methods. The only parts that actually depend on the measurement scheme are the calculation of $T_{mn}$ and of one gradient (see \eqref{eq:gradient_step}). For collinear schemes only the expression for $\Hf(\delta_m, \wb_n)$ has to be replaced.

Convergence on the global minimum is not guaranteed with COPRA.
In principle, it has to be restarted repeatedly from different initial values. However, it was designed to achieve high retrieval probabilities even for totally random initial guesses. With an informed initial guess usually a single run of COPRA is sufficient~(see Sec.~\ref{subsec:copra}).

In a setup step the trace error $R$ and the scaling factor $\mu$ from \eqref{eq:scaling_factor} are calculated for the initial guess. Furthermore, the maximum local gradient norm $g^0_{M-1}$ (see next section) has to be calculated here before the algorithm starts with the first of two stages.

\subsection{Stage I: local iteration}
In the first stage all steps are performed subsequently on one spectrum at a time, i.e., $m$ is constant below. Hence, we call this stage local iteration. The spectra are processed in random order, but for notations sake we assume that we start with $m=0$ and end with $M-1$. The corresponding one-dimensional measurement signal $S_{mk}$ and its spectrum $\tilde{S}_{mk}$ are calculated by Eqs.~(\ref{eq:complex_measurement_trace}-\ref{eq:complex_measurement_spectrum}). Next a projection on the measured intensity is performed. For that the amplitude of $\tilde{S}_{mk}$ is replaced by the measured one from $\Tm_{mn}$, followed by an inverse Fourier transform to obtain a new measurement signal $S_{mk}'$:
\begin{align}
 S_{mk}' = \mu \: \idft{k}{n} \left( \tilde{S}_{mn} / | \tilde{S}_{mn} | \sqrt{T^\mathrm{meas}_{mn}}\right). \label{eq:projection_step}
\end{align}
Note the use of the scaling factor $\mu$ here. We define the distance $Z_m$ between $S_{mk}'$ and $S_{mk}$ as
\begin{align}
Z_m = \sum_{k} \left|S_{mk}' - S_{mk}(\vE) \right|^2,
\end{align}
and try to minimize it in terms of the current solution $\vE$. To that end, in iteration $j$ a single gradient descent step is performed for every spectrum $m$
\begin{align}
 \senv_n' = \senv_n - \gamma_m^j \: \nabla_n Z_m. \label{eq:gradient_step}
\end{align}
The expressions for the gradient $\nabla_n Z$ for all PNPS methods discussed here are given in the supplementary material~(Sec. S3). They can be evaluated using one or two additional fast Fourier transforms~(FFT) depending on the scheme.

The algorithm proceeds to the next spectrum using the updated solution $\vE'$. If all spectra are processed one local iteration of the algorithm is finished. Additionally, after every iteration $R$ is calculated from $S_{mk}$ and $\mu$ is updated. The local iteration is stopped when no improvement of $R$ was achieved for ten iterations.

The step size $\gamma$ is crucial to the convergence of the local iteration. For traces without or with very little noise we found the following to work very well
\begin{align}
  \gamma = Z_m / \sum_{n} |\nabla_n Z_m|^2. \label{eq:noiseless_step}
\end{align}
In the presence of noise, however, this choice leads to poor convergence. We found that reliable convergence for all noise levels can be achieved by exchanging the denominator. For that we keep track of the maximum gradient norm in every iteration~$j$
\begin{align}
g^j_m = \mathrm{max} \bigr(g^j_{m-1}, \sum_{n} |\nabla_n Z_m|^2\bigl) \quad \text{with} \quad g^j_{-1} = 0. \label{eq:gradient_norm}
\end{align}
The step size in iteration $j$ is then defined by
\begin{align}
\gamma_m^j = Z_m / \mathrm{max}(g^{j}_m, g^{j-1}_{M-1}),
\end{align}
where $g^{j}_m$ is the running estimate for the maximum gradient norm in the current iteration and $g^{j-1}_{M-1}$ is the maximum gradient norm encountered during the last iteration. Before the first local iteration $g^{0}_{M-1}$ has to be determined separately in the setup step which counts as the first iteration $j=0$.

\subsection{Stage II: global iteration}
In the second stage all spectra are processed simultaneously in every step and $m$ runs from $0$ to $M-1$ in the expressions below. Hence, we call this stage global iteration. It is seeded by the best solution of the local iteration stage.

A global iteration starts by calculating $S_{mk}$, $\tilde{S}_{mn}$, and $T_{mn}$ for the current guess $\vE$ by Eqs.~(\ref{eq:complex_measurement_trace}-\ref{eq:measurement_trace}). Then the trace error $R$ and the scale factor $\mu$ are computed by Eqs.~(\ref{eq:data_error}-\ref{eq:scaling_factor}). An updated signal $S_{mk}'$ is obtained by minimizing $r$ from \eqref{eq:residuals} in terms of $S_{mk}$. This is done by a single gradient descent step
\begin{align}
  S_{mk}' &= S_{mk} - \eta_r \: \nabla_{mk} \nnR , \label{eq:m_step} \\
  \shortintertext{with}
  \eta_r &= \alpha \bigl( \nnR  / \sum_{lj} | \nabla_{lj} \nnR |^2 \bigr), \label{eq:gamma1} \\
  \shortintertext{where the gradient is given by}
  \nabla_{mk} \nnR  = -4 \mu \frac{\Delta t}{2\pi \Delta \omega}& \idft{k}{n} \left[ \bigl( \Tm_{mn} - \mu T_{mn} \bigr) \tilde{S}_{mn} \right]. \label{eq:glob_gradient}
\end{align}
We follow up by adapting $\vE$ to this new estimate $S_{mk}'$. This is done as in the local iteration, but all spectra are processed simultaneously. With
\begin{align}
Z = \sum_{m} Z_m = \sum_{mk} \left|S_{mk}' - S_{mk}(\vE) \right|^2,
\end{align}
we have
\begin{align}
\nabla_n Z = \sum_{m} \nabla_n Z_m.
\end{align}
We obtain the next estimate by a single gradient descent step
\begin{align}
\senv_n' &= \senv_n - \eta_z \nabla_n Z, \label{eq:global_step} \\
\shortintertext{with}
\eta_z &= \alpha \bigl( Z / \sum_k | \nabla_k Z |^2 \bigr). \label{eq:gamma2}
\end{align}
The constant $\alpha$ controls the step size both in \eqref{eq:gamma1} and \eqref{eq:gamma2}. We use $\alpha=0.25$ for all results shown in this work.

In this work we simply performed COPRA for a fixed number of total iterations. However, an arbitrary convergence criterion can be used instead to terminate the global iteration. After the algorithm has terminated, the solution with the lowest trace error $R$ is returned.

\subsection{Design considerations}
Fig.~\ref{fig:algorithm} summarizes the discrete PNPS formalism and shows diagrams of both stages of the algorithm. In the following we will discuss the overall design and implementation of the algorithm and how it relates to existing approaches.

The local iteration aims to provide an approximation of the solution in a rapid and reliable way. It is less likely to get stuck in a local minimum and the initial convergence is much faster than during the global iteration. For the noiseless case, i.e., on synthetic measurements, only this stage of COPRA is necessary. However, in the presence of noise it will fail to converge to a least-squares solution. This is why the global iteration has to be performed subsequently.

Our algorithm was inspired by GPA for FROG~\cite{DeLong1994,Trebino2002}. For example, we minimize the same distance $Z$ in our algorithm. However, there are some key differences. 

First of all, COPRA operates with the pulse spectrum $\vE$ as the independent variable. Choosing the pulse field $\mathbf{E}$ would in general make the calculation of the PNPS trace and the required gradients more complicated.

Second, we found heuristically safe and divergence-free expressions for the step sizes. This avoids the overhead of determining them with a line search in every iteration.

Third, the local iteration processes one spectrum at a time.
We found that this approach can increase the convergence speed and makes the algorithm less prone to stagnation.

Finally, the most important difference is that in the global iteration we replaced the projection on the measured intensity by a gradient descent step.
This it what allows to obtain a least-squares solution which is not possible when using a projection on the measurement.

The local iteration is similar to ptychography-based algorithms for SHG-TDP~\cite{Spangenberg2015} and SHG-FROG~\cite{Sidorenko2016,Sidorenko2017}, since the update step in ptychography is a gradient descent step~\cite{Maiden2017}. However, the gradients used in ptychography are different, as the expression for SHG is seen as linear in two independent variables: the object and the probe pulse. Consequently, the gradient in ptychography is calculated with respect to one pulse only. This is only one of two terms used in our algorithm. This issue is discussed in detail in the supplementary material~(Sec.~S6).

\section{Methods}\label{sec:methods}
For testing purposes we created an overall number of 100 random test pulses with a root-mean-square time-bandwidth product~(TBP) of $2$. For comparison, the TBP of a Gaussian pulse with flat phase is $0.5$ in this definition. The pulses possess a complex amplitude and phase structure in both the time and frequency domain. The grid size was $N=256$. Retrieving such pulses represents a significant challenge for pulse retrieval algorithms and allows us to clearly assess the performance of our algorithm.

The specific central frequency ($\lambda_0 = \SI{800}{\nm}$) and the temporal grid spacing ($\Delta t= \SI{5}{\fs}$) used in our simulations have no influence on the pulse retrieval. The results obtained here are applicable to other frequency and time scales, thus, we leave out the information on the frequency and time axes.

The algorithm was usually initialized by a Gaussian pulse with a duration of $\SI{50}{\fs}$~(full width half maximum) and random spectral phase (uniformly distributed on $[-0.1\pi, 0.1\pi]$). This matches realistic conditions where only rough knowledge about the pulse duration is available. 

To test COPRA under stricter conditions we also used a completely random initial guess. Its spectral amplitude and phase were uniformly distributed on $[0, 1]$ and $[0, 2\pi]$ respectively. This makes no assumptions at all and allows for a fully unbiased estimation of the retrieval probability of the algorithm.

We quantified the retrieval accuracy by comparing the retrieved solution $\vE$ to the test pulse $\vE^0$ from which the synthetic measurement trace was generated. This was done by calculating the \emph{retrieval error} $\varepsilon$, which is the NRMSE between both pulses. However, the ambiguity of the constant and linear spectral phase as well as the scaling have to be taken into account. This leads to the formal definition
\begin{align}
\varepsilon(\vE) &\equiv \Bigl[\: \min_{\mu, \varphi_0, \varphi_1} \sum_n |\senv_n^0 - \mu \exp[\im (\varphi_0 + \varphi_1 \wb)] \senv_n|^2  \nonumber \\
& \qquad / (N \: \max_{n} |\senv_n^0|^2) \Bigr]^{1/2}. \label{eq:pulse_error}
\end{align}
Additionally, for some schemes the time-reversal ambiguity has to be considered, in which case $\varepsilon \equiv \min[\varepsilon(\vE), \varepsilon(\vE^*)]$. The procedure of how $\varepsilon$ is calculated is described in the supplementary material~(Sec.~S4).

To investigate the influence of noise on the retrieval we added Gaussian noise to the synthetic measurement traces. The standard deviation $\sigma$ of the noise was chosen relative to the maximum intensity of the trace and is given in percent, e.g., $\sigma = 1\%$. This noise model corresponds to a low intensity measurement with a CCD array spectrometer where signal-independent noise sources dominate~\cite{Davenport2015}. Signal-dependent Gaussian noise requires to introduce a weighting in the pulse retrieval problem, which leads to small modifications in COPRA as is discussed in the supplementary material~(Sec.~S8).

In the noiseless case $R$ directly quantifies the convergence and is only limited by the accuracy of the trace computation. We assumed successful retrieval if a solution with $R < \num{1e-4}$ was obtained. However, for noisy measurements $R$ will be on the order of the relative noise level, i.e., $\sigma = 1\%$ leads to $R \approx 1\%$. To assess the convergence we compare it to the non-vanishing trace error of the test pulse $\vE^0$ used to create the synthetic measurement
\begin{align}
 R_0 \equiv R(\vE^0). \label{eq:min_trace_error}
\end{align}
This is an approximation of the expected trace error of the global least-squares solution. Specifically, we assume successful retrieval if $R < R_0 + \num{1e-4}$. 

To assess the versatility of COPRA we tested it on a multitude of PNPS schemes, including common ones such as SHG-FROG, PG-FROG, SHG-TDP, SHG-d-scan, SHG-iFROG, as well as less common ones such as THG-d-scan, SD-d-scan and THG-iFROG. Furthermore, we included variants of MIIPS and iFROG, namely THG-MIIPS, SD-MIIPS and SD-iFROG, that to our knowledge have not yet been demonstrated experimentally. They serve to showcase COPRA's universality.

For every scheme we selected an appropriate parameter set $\delta_m$. For FROG methods we chose to sample the delay $\tau$ like the pulse itself with $\tau_m = t_m$ and $M=N$. This is the common choice and the one required by PCGPA. For iFROG the same choice was used except when using SD as the nonlinear process. In this case we sampled $\tau$ at four times the frequency with $\Delta \tau = 0.25 \Delta t$ and $M = 4N$. For the other schemes we sampled $\delta_m$ with $M=128$ points. For MIIPS the free parameters $\alpha$ and $\gamma$ had to be chosen appropriately. Details and the full list of parameters can be found in the supplementary material~(Sec.~S4).

For SHG-FROG we compare our algorithm to two other fast pulse retrieval algorithms: PCGPA and a recently proposed retrieval algorithm based on the ptychographic iterative engine~(PIE)~\cite{Sidorenko2016,Sidorenko2017}. We give details on this choice and their implementation in the supplementary material~(Sec.~S6).

\section{Results}\label{sec:results}

\subsection{Nonlinear least-squares solvers}\label{subsec:nonlinear_solvers}
\begin{figure}[tb]
	\centering
	\includegraphics{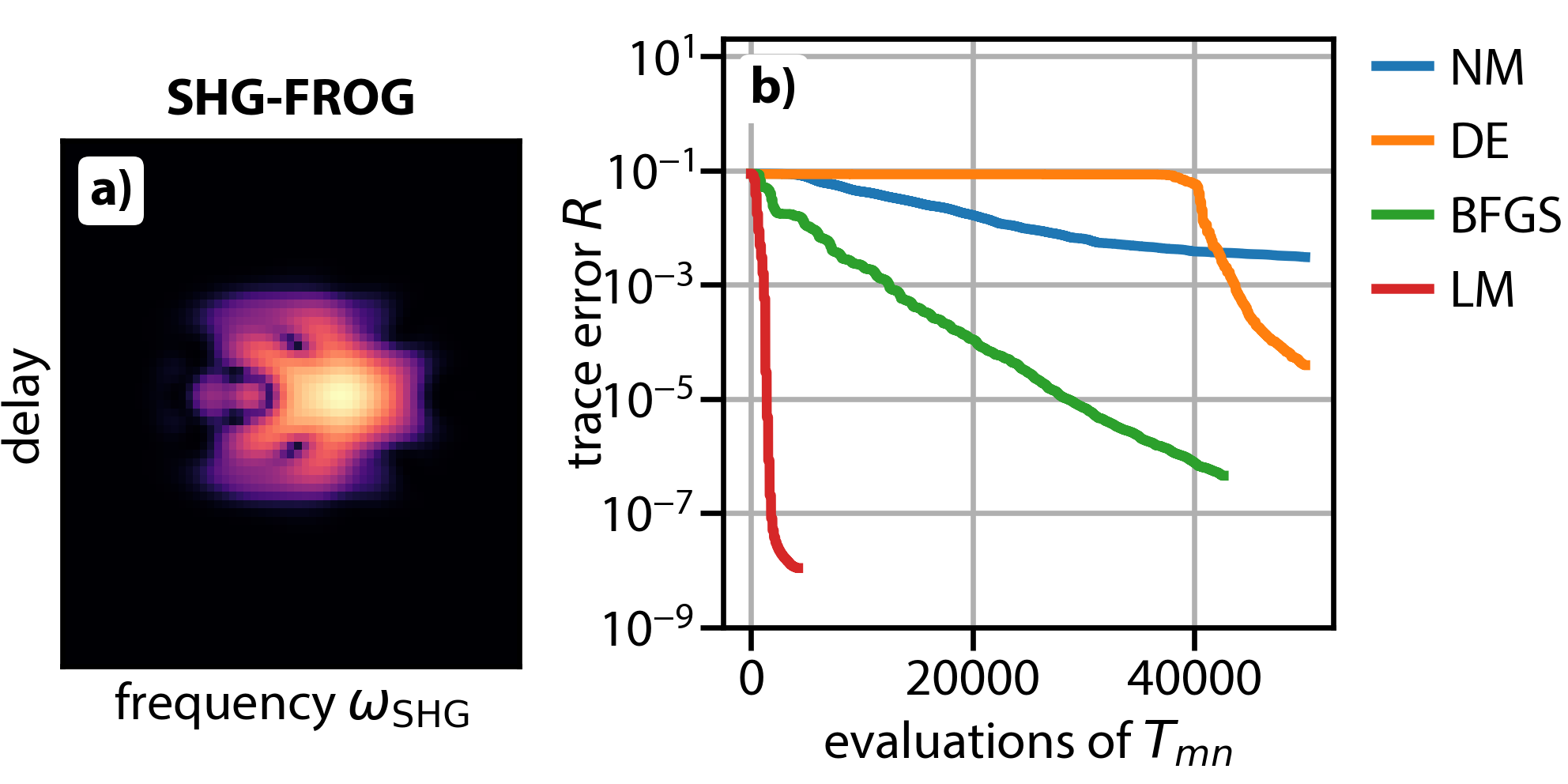}
	\caption{a) Synthetic SHG-FROG trace with $N=M=64$. b) Trace error obtained by four different minimization algorithms plotted over the number of full evaluations of the trace $T_{mn}$. Shown are the best out of ten runs for every algorithm.}
	\label{fig:levenberg}
\end{figure}
To demonstrate the applicability of general minimization algorithms to the pulse retrieval problem we created a simple SHG-FROG trace with $N=M=64$ belonging to a pulse with TBP~1~(see Fig.~\ref{fig:levenberg} a). Then we retrieved pulses by using four different minimization algorithms: Nelder-Mead~(NM) and differential evolution~(DE) are scalar, gradient-free minimization methods used to retrieve pulses from d-scan and iFROG measurements~\cite{Miranda2012,Hyyti2017b,Escoto2018}. Broyden–Fletcher–Goldfarb–Shanno~(BFGS) is a scalar, gradient-based algorithm, which was used to retrieve pulses from chirp scan and FROG measurements~\cite{Wilcox2014}. Furthermore, we tested the Levenberg-Marquadt~(LM) algorithm, which is a specialized nonlinear least-squares solver. For both BFGS and LM numerical differentiation was used to approximate the derivatives. The additional trace evaluations required for that were factored into our comparison.

The convergence behavior in terms of trace evaluations for the best of the ten retrievals from random initial guesses is shown in Fig.~\ref{fig:levenberg}~b). LM massively outperformed the other algorithms in terms of retrieval efficiency. For every run a solution ($R < \num{1e-4}$) was found with an effort of less than $1500$ trace evaluations. NM and DE on the other hand showed slow convergence. This result is reasonable. Since $T(\omega, \delta; \vE)$ is smooth and differentiable in $\vE$ gradient-based algorithms are favored.

This demonstrates that, in theory, there is no need for a specialized pulse retrieval algorithm for PNPS measurements. However, the run time of the LM approach scales badly~\cite{Wright1999}. In practice it becomes infeasible when retrieving complex pulses that require large simulation grid sizes ($N > 256$) and many spectral measurements ($M \gtrsim N$). Retrieving a pulse may then take several hours on a normal workstation compared to the tens of seconds required for the measurement in Fig.~\ref{fig:levenberg}.

More details can be found in the supplementary material~(Sec.~S4).

\begin{figure*}[tb]
\centering
\includegraphics{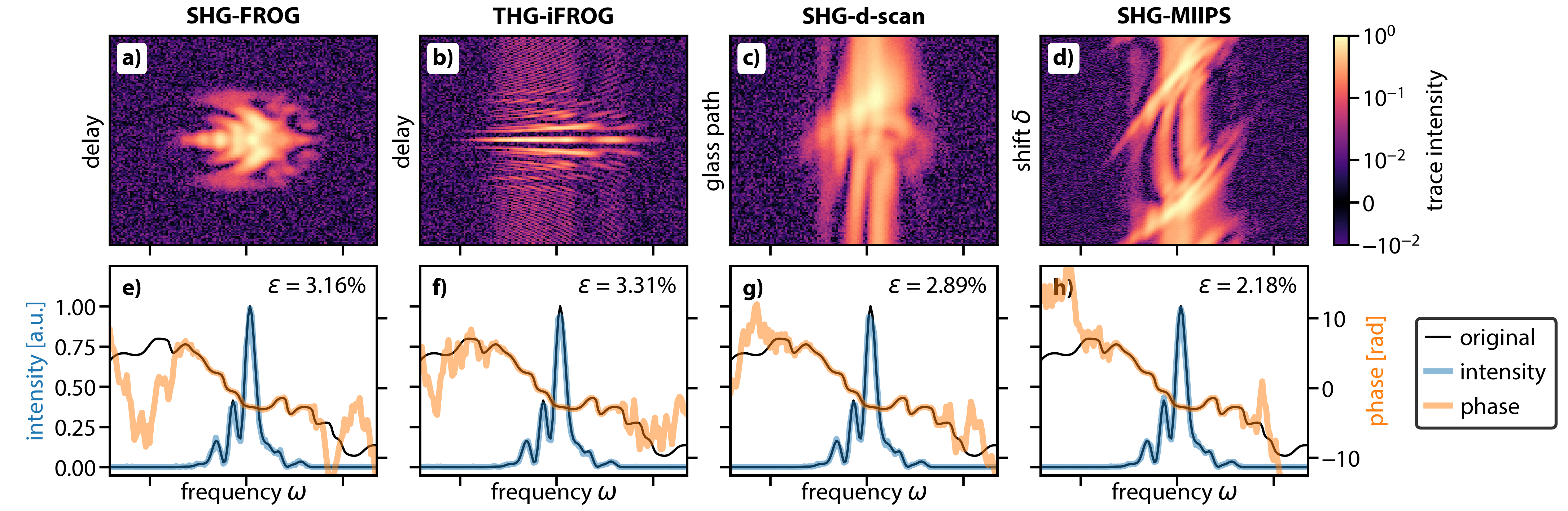}
\caption{Pulse retrieval from different PNPS measurements using COPRA: a)-d) Synthetic measurement traces with added Gaussian noise ($\sigma = 1\%$). e)-h) The retrieved pulses~(blue: intensity, orange: phase) and the original pulse (in black). The retrieval error $\varepsilon$ quantifies the retrieval accuracy. The test pulse has a time-bandwidth product of 2.}
\label{fig:show_noisy_pulse_retrieval}
\end{figure*}

\subsection{COPRA}\label{subsec:copra}
To assess the performance of COPRA we ran a large pulse retrieval simulation on synthetic PNPS measurements. We performed $10$~runs of COPRA for all $100$~test pulses for $7$ noise levels ($\sigma=$ $0\%$, $0.1\%$, $0.3\%$, $0.5\%$, $1\%$, $3\%$, $5\%$) for all PNPS schemes. For $\sigma=0\%$ only the local stage of COPRA with the step size from \eqref{eq:noiseless_step} was used. In all cases 300~iterations were performed. The retrieval was initialized with a Gaussian pulse with random phase~(see Sec.~\ref{sec:methods}).

For illustration, four measurement traces with $\sigma=1\%$ and the pulses retrieved from them are shown in Fig.~\ref{fig:show_noisy_pulse_retrieval}. In the following we will discuss the results in detail.

\paragraph{Convergence speed}
\begin{figure}[tb]
\centering
\includegraphics{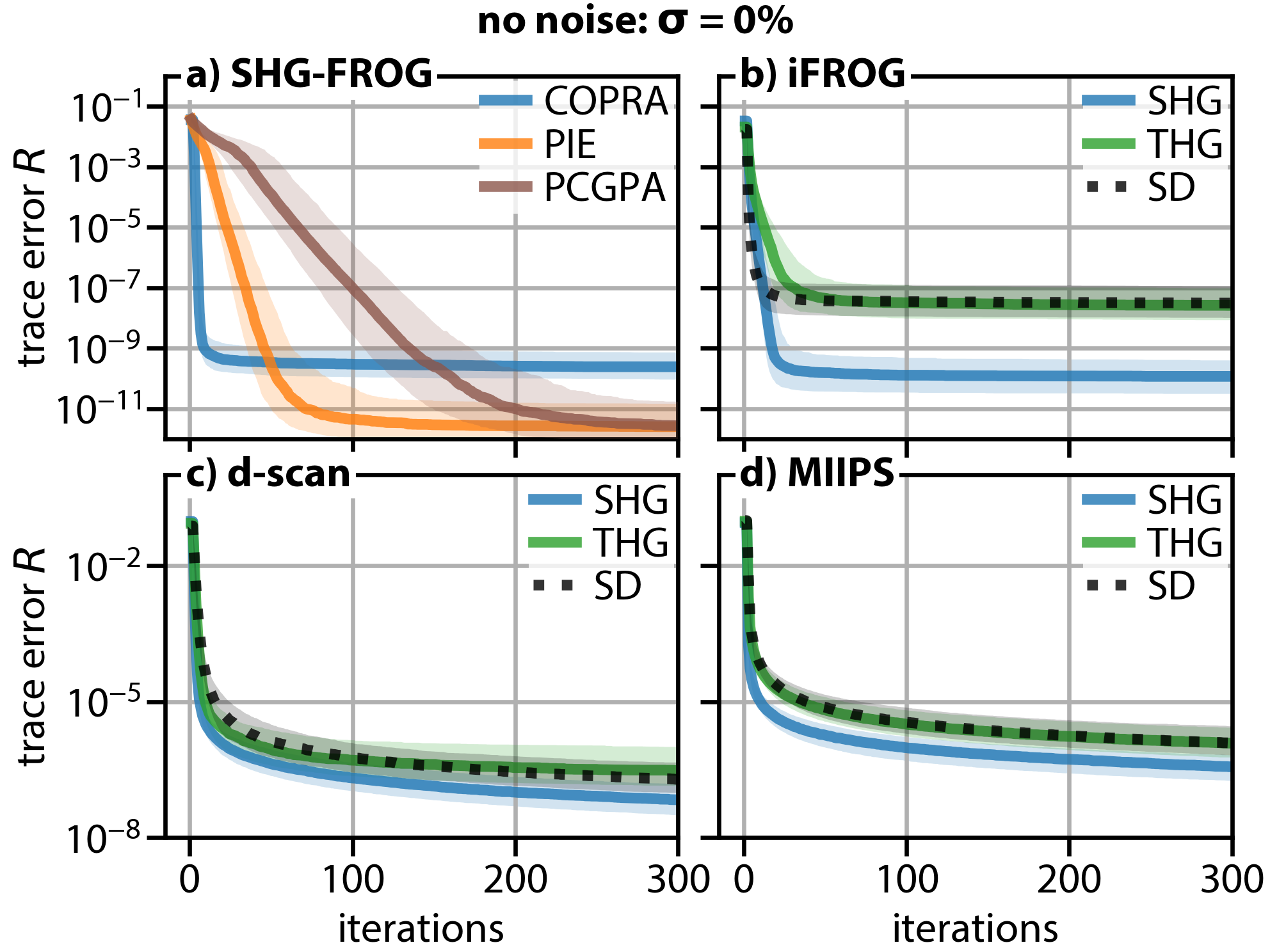}
\caption{Convergence behavior for $\sigma=0\%$. a) comparison of algorithms, b-d) comparison of PNPS methods when retrieving with COPRA. Shown is the median (bold line) and the interquartile range (shaded area) of the running minimum of $R$. Only the local iteration was used in COPRA.}
\label{fig:convergence_behavior}
\end{figure}
\begin{figure}[tb]
\centering
\includegraphics{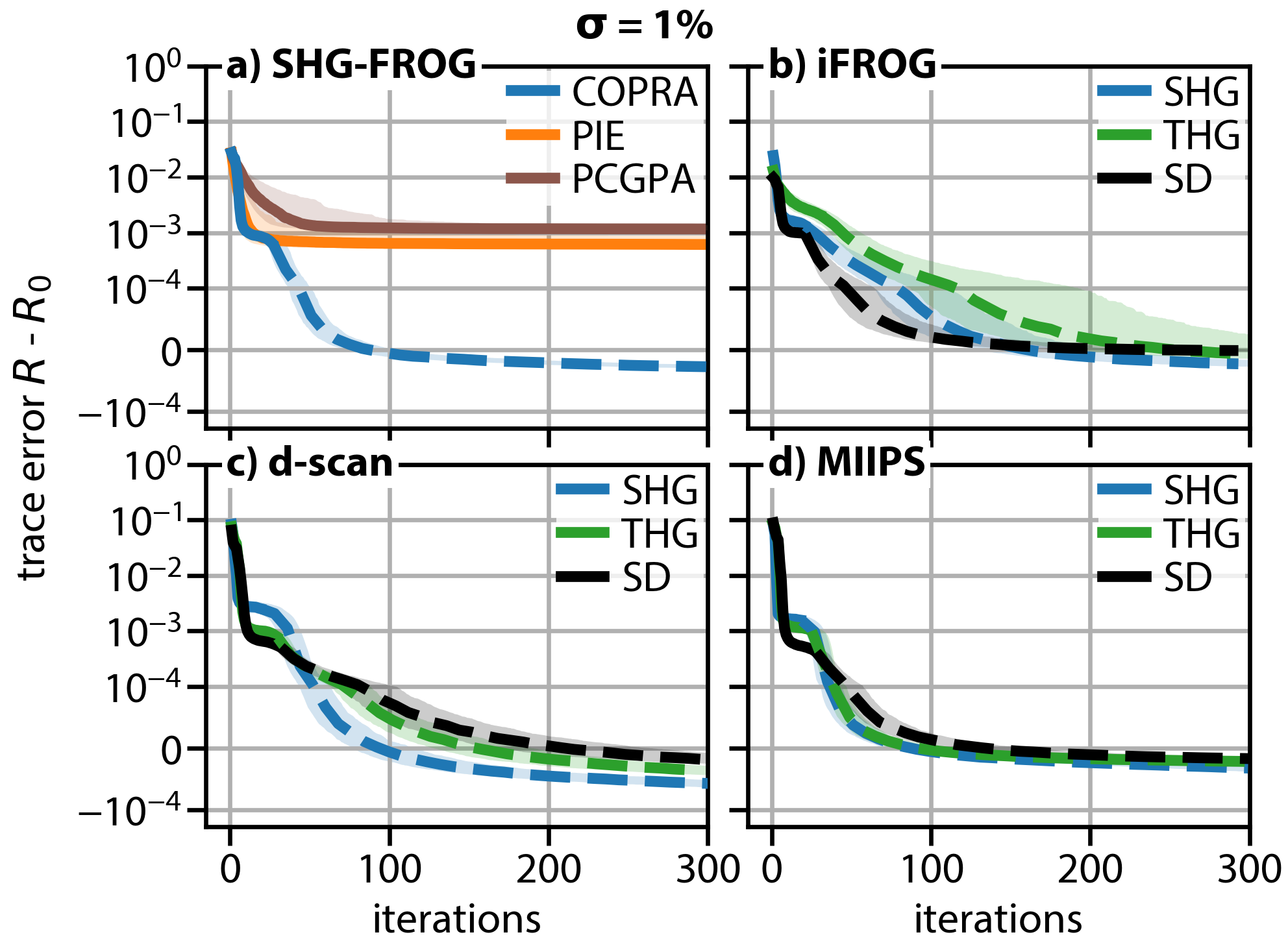}
\caption{Convergence behavior for noisy measurements~($\sigma = 1\%$). a) comparison of algorithms, b-d) comparison of PNPS methods when retrieving with COPRA (solid: local iteration, dashed: global iteration). Shown is the median (bold line) and the interquartile range (shaded area) of the running minimum of $R$.}
\label{fig:noisy_convergence_behavior}
\end{figure}
The typical convergence behavior of the local iteration for noiseless traces is shown in Fig.~\ref{fig:convergence_behavior}. It is important to note that the figures show the running minimum of the trace error, i.e., the best solution encountered. COPRA does not reduce the trace error in every step. For non-collinear schemes such as SHG-FROG we observed convergence to the accuracy limit of COPRA ($R \sim \num{1e-9}$ for SHG) within just $20$~iterations, which is less than PCGPA and PIE require. For other PNPS schemes the initial convergence is just as fast, reaching the threshold of $R < \num{1e-4}$ usually within tens of iterations. 

For d-scan and MIIPS, especially for the variants using third-order nonlinearities, the convergence slows down at some point. We attribute this to the conditioning of the problem, which is known to directly affect the convergence speed of gradient descent methods~\cite{Wright1999}. However, usually this has no impact as stagnation sets in only below $R \sim \num{1e-5}$~(see also Sec.~\ref{par:uniqueness}).

The typical convergence behavior for noisy measurements is shown in Fig.~\ref{fig:noisy_convergence_behavior}. To quantify the convergence to the least-squares solution in this case the difference between the trace error $R$ and the trace error of the test pulse $R_0$ from \eqref{eq:min_trace_error} is used. The results demonstrate the role of the two stages in COPRA. The local iteration converges rapidly and then stagnates at roughly $R_0 + \num{1e-3}$. Afterwards, it is the global iteration that continues to minimize the squared sum of residuals $r$ and actually solves the pulse retrieval problem for noisy data in the least-squares sense.

In practice, COPRA converges very fast. For many PNPS measurements it finds a good solution with less than $100$ within iterations. For other PNPS methods a few hundred iterations are usually sufficient. Also, its actual run time compares favorably with other fast retrieval algorithms such as PCGPA and PIE and massively outperforms general minimization algorithms such as LM. For example, a single SHG-FROG retrieval with COPRA ($100$~iterations) on a grid with $N=256$ takes less than $\SI{3}{\s}$~sec on a normal workstation. Furthermore, a single COBRA iteration requires only $5M$ to $7M$ 1D-FFTs with $N$ elements (depending on the scheme and the stage) and no operations with higher computational complexity. Hence, its computational complexity is approximately $MN\log N$.

\paragraph{Retrieval accuracy and retrieval probability}
\begin{figure*}[tb]
    \centering
    \includegraphics{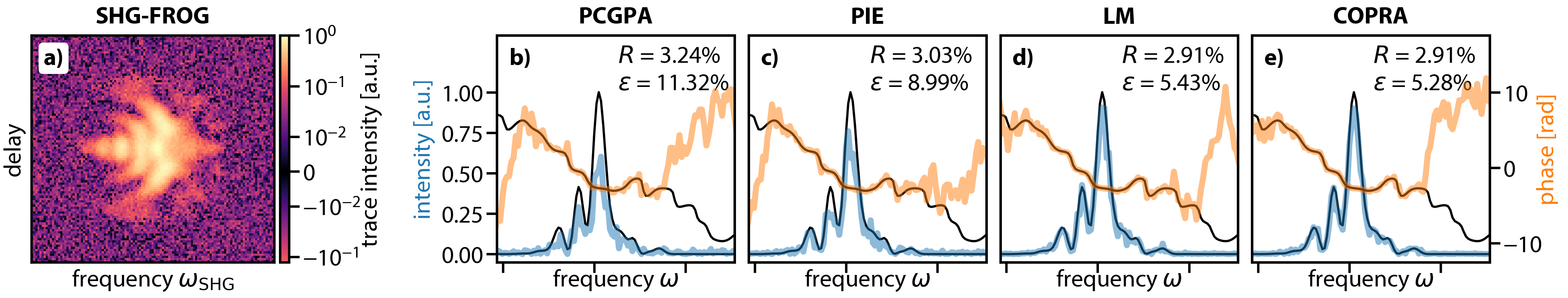}
    \caption{Robustness against additive Gaussian noise: a) synthetic SHG-FROG trace of a pulse with TBP 2. A high level of Gaussian noise was added~($\sigma = 3\%$). b)-e) The pulses retrieved from the trace using different algorithms (blue: intensity, orange: phase) compared to the test pulse (black). $\varepsilon$ quantifies the retrieval accuracy.}
    \label{fig:show_robust_pulse_retrieval}
\end{figure*}
\begin{figure}[tb]
    \centering
    \includegraphics{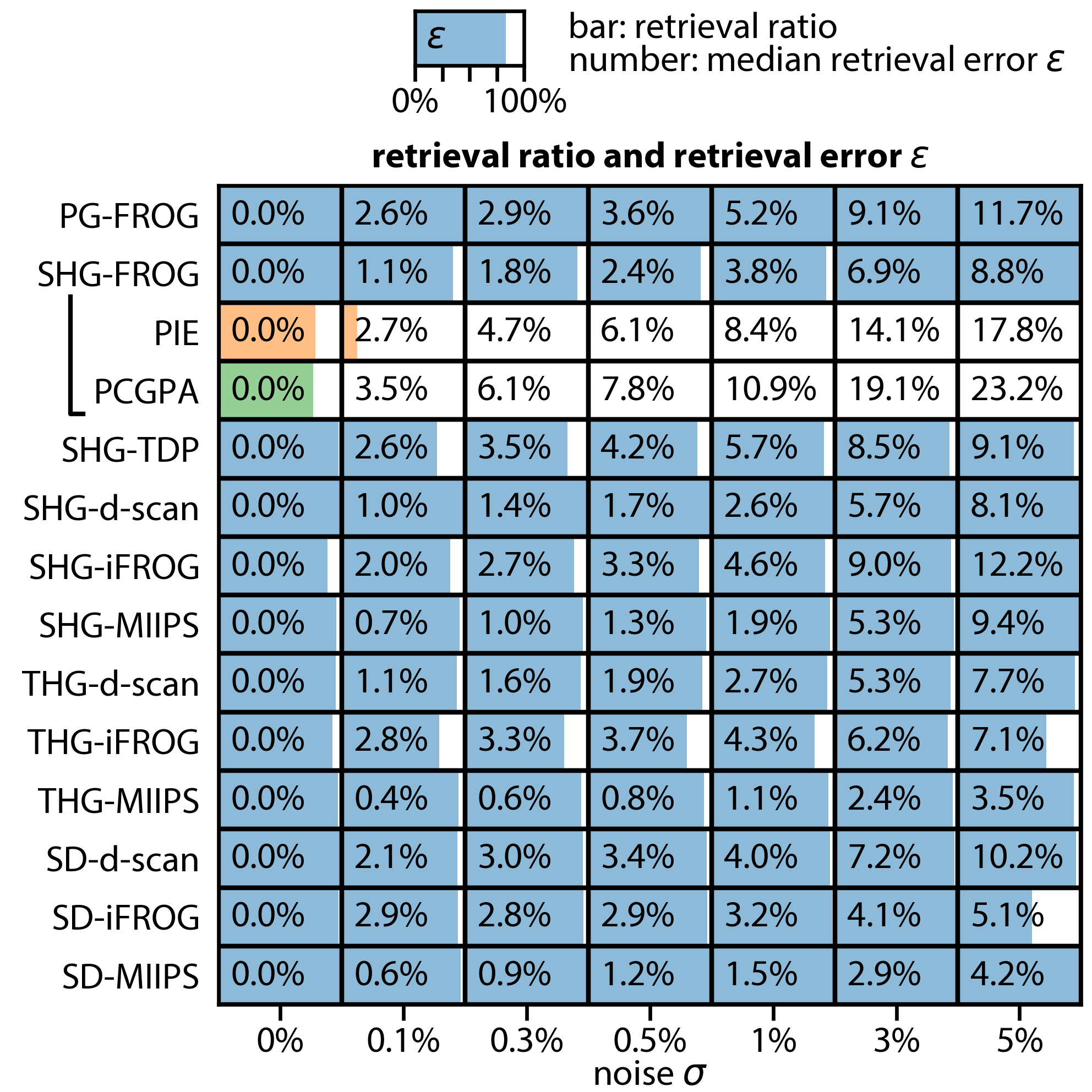}
    \caption{The retrieval ratio (blue bar) and median retrieval error (inset number) of COPRA in dependence of the PNPS method and the noise level. A comparison with PCGPA and PIE for SHG-FROG is included. Successful retrieval was assumed if $R < R_0 + \num{1e-4}$.}
    \label{fig:show_pulse_retrieval}
\end{figure}
In Fig.~\ref{fig:noisy_convergence_behavior} we see that PCGPA and PIE \emph{do not} achieve convergence in the least-squares sense. We found that, in general, no algorithm that incorporates the measurement trace solely by a projection has this property. This includes every algorithm based on generalized projections or the ptychographic engine. Both methods stagnate at roughly $R \sim R_0 + \num{1e-3}$ depending on the noise level -- similar to the local iteration stage of COPRA.

The impact of this can be seen in Fig.~\ref{fig:show_robust_pulse_retrieval}. It shows a comparison of the pulses retrieved from a very noisy SHG-FROG trace ($\sigma = 3\%$) by using PCGPA, PIE, LM and COPRA. In each case the best solution after 10~runs of the algorithm is shown. The solutions obtained by PCGPA and PIE are clearly less accurate, having a retrieval error of $\epsilon=11.32\%$ and $\epsilon=8.99\%$ compared to $\epsilon=5.43\%$ for COPRA. The pulse retrieved by LM confirms that COPRA does, in fact, obtain the least-squares solutions. At the same time one run of LM took $\SI{272}{\s}$ to complete, compared to $\SI{7}{\s}$ for COPRA.

In Fig.~\ref{fig:show_pulse_retrieval} we show the retrieval ratio, i.e., the percentage of retrieved solutions that fulfill $R < R_0 + \num{1e-4}$, for all combinations of noise levels and PNPS schemes. The retrieval probability of COPRA is very high in all cases and usually above $90\%$. In practice a few repeated runs of COPRA from different initial guesses suffice. Also, we see that virtually none of the solutions obtained by PCGPA and PIE for noisy measurements fulfill our convergence criterion.

Additionally, Fig.~\ref{fig:show_pulse_retrieval} shows the retrieval errors achieved in the different cases. Shown is the median of the minimum retrieval error from $10$ runs achieved for each of the $100$~synthetic measurements. The results for SHG-FROG also show that PCGPA and PIE are less accurate than COPRA for all noise levels $\sigma>0\%$. This confirms the discussion from above. 

We found that some PNPS schemes are more sensitive to noise and more susceptible to lack of convergence. E.g., SHG-iFROG measurements with $\sigma=1\%$ lead to $\epsilon=4.5\%$ compared to $\epsilon=2.6\%$ for SHG-d-scan. However, the actual dependence of $\epsilon$ on the PNPS scheme, the trace error and the noise level is complex and a full description is out of scope for this work.

\paragraph{Uniqueness of the retrieved solutions}\label{par:uniqueness}
To verify the feasibility of the PNPS schemes as full pulse measurement methods we tested the uniqueness of the pulses retrieved by COPRA from noiseless, synthetic PNPS traces. To increase the search range we repeated the retrieval simulation starting COPRA from a random initial guess. Specifically, we searched for solutions of the retrieval problem that have a small trace error and simultaneously a large retrieval error, i.e., $R < \num{1e-4}$ and $\epsilon > 1\%$. This would indicate the existence of a non-trivial ambiguity in one of these methods. To verify these solutions they were refined to high accuracy using the LM algorithm.

In this study we found no occurrence of an exact non-trivial ambiguity for any of the tested PNPS schemes. This is an indication that these measurements, in theory, define amplitude and phase of the pulse uniquely up to the trivial ambiguities. This includes MIIPS measurements, which to our knowledge have only been used for pulse compression so far.

However, we found that pulse retrieval from d-scan and MIIPS measurements may admit solutions with very low trace errors ($R \approx \num{1e-5}$) that have additional weak satellite pulses at large delays. As those disappear after further refining of the solution to a level of $R < \num{1e-9}$ the solutions do not constitute an ambiguity of the scheme in the strict sense. However, they will impact the retrieval from real, noisy measurements. This indicates that in general pulse retrieval from PNPS measurements should use some kind of regularization to select the correct solution. This can be done implicitly by starting COPRA with a temporally localized pulse, e.g., a Gaussian in the time domain. The satellite pulses only appeared as solutions for random initial guesses.

Notably COPRA performed well for many PNPS schemes even when using uninformed, random initial guesses. The retrieval ratio was mainly impacted for MIIPS measurements which admit several local solutions due to the periodicity of the applied phase patterns. The full results of the second retrieval simulation and further discussion of the satellite pulses in d-scan and MIIPS can be found in the supplementary material~(Sec. S7).

\subsection{Spectrally incomplete traces}\label{subsec:partial_traces}
\begin{figure}[tb]
\centering
\includegraphics{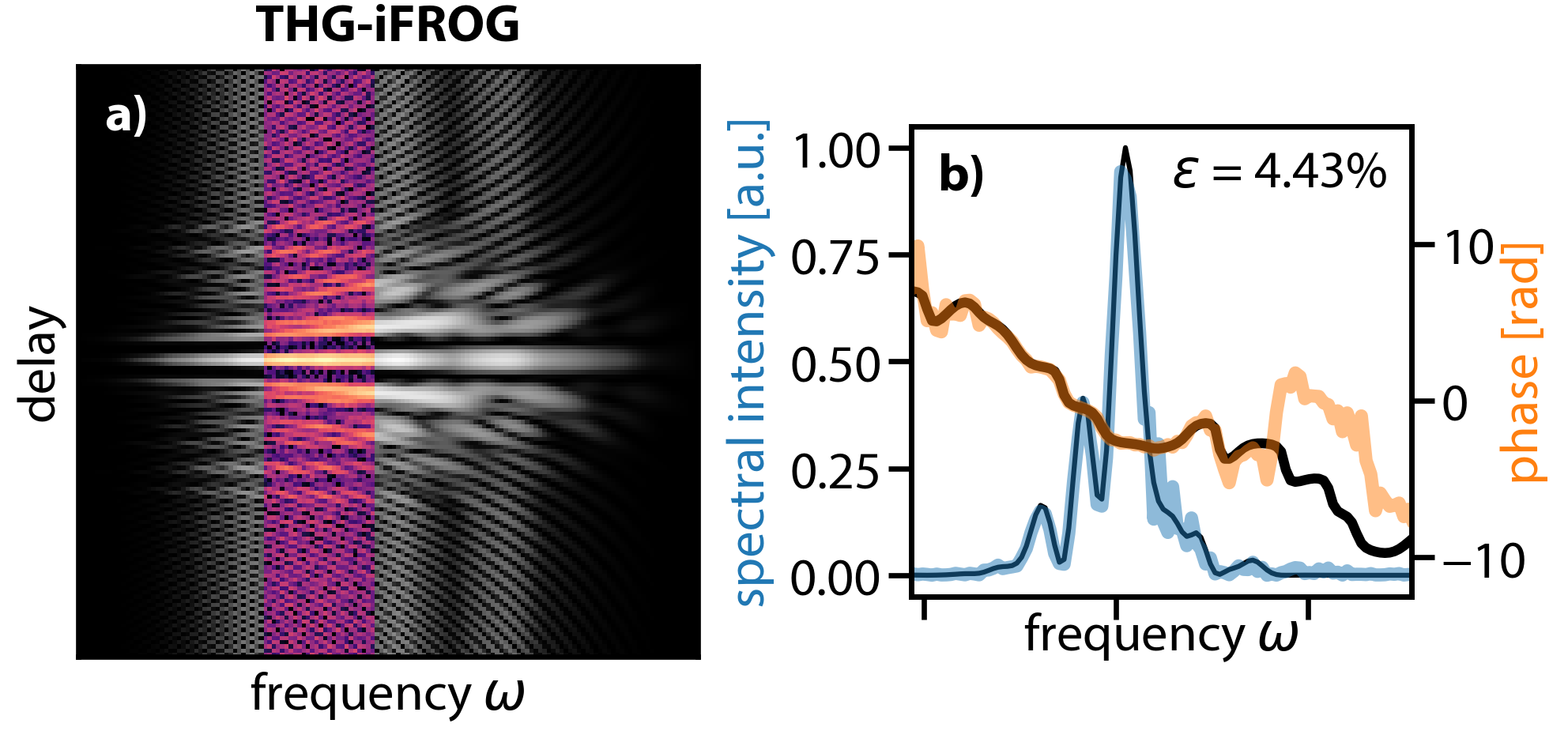}
\caption{Pulse retrieval from an incomplete THG-iFROG measurement. a) Synthetic measurement trace with added noise ($\sigma = 1\%$). $25$ of $N=256$ spectral measurements were used for retrieval (indicated by color) b) The retrieved pulse~(blue: intensity, orange: phase) and the test pulse (black). $\varepsilon$ quantifies the retrieval accuracy.}
\label{fig:show_partial_pulse_retrieval}
\end{figure}
Sometimes it may be required to retrieve pulses from spectrally incomplete measurement traces. This may be due to, e.g., overlap with the fundamental spectrum or limitations of the spectrometer. The retrieval from spectrally incomplete traces was already demonstrated for d-scan and FROG~\cite{Miranda2012a,Sidorenko2016}.

With small modifications COPRA can work with such traces. Mainly, Eqs.~(\ref{eq:projection_step}) and (\ref{eq:glob_gradient}) have to be changed to include only the available $\omega_n$. Using this version of COPRA we found that retrieval from spectrally incomplete traces is possible for all PNPS schemes. Fig.~\ref{fig:show_partial_pulse_retrieval} shows an example of an incomplete THG-iFROG measurement for which to our knowledge retrieval has not yet been demonstrated. The noisy measurement trace ($\sigma = 1\%$) is the same as in Fig.~\ref{fig:show_noisy_pulse_retrieval} b) except that less than $10\%$ of the frequency range was selected~(only $25$ spectral sampling points of $N=256$). Additionally, we included values of zero intensity in the measurement trace spanning the first and last $15\%$ of the simulation grid in frequency direction~(not in the range shown in Fig.~\ref{fig:show_partial_pulse_retrieval}) which improved the retrieval accuracy by enforcing the localization of the pulse in the frequency domain. We can see that retrieval is possible with only a moderate loss of retrieval accuracy, i.e., the retrieval error is $\varepsilon=4.43\%$ compared to $\varepsilon=3.31\%$ for the complete trace from Fig.~\ref{fig:show_noisy_pulse_retrieval} b).

\section{Conclusion}\label{sec:conclusion}
In conclusion, we showed that many self-referenced pulse measurement schemes are conceptually similar and can be described within a common mathematical framework. They measure the same quantity: sets of parametrized nonlinear process spectra~(PNPS).

The PNPS pulse retrieval problem is naturally formulated as a nonlinear least-squares problem. Its solution is a maximum-likelihood estimate under the experimentally relevant assumption of Gaussian noise. This aspect was not fully appreciated before and methods that project on the measured intensity such as generalized projections and ptychography do not obtain a least-squares solution. Consequently, the accuracy of the retrieved solutions suffers unnecessarily in the presence of measurement noise.

The main result of the paper is the common pulse retrieval algorithm~(COPRA), which can be directly applied to all PNPS measurements. We verified and demonstrated its capabilities numerically, by algorithmic testing of a large suite of synthetic PNPS traces generated from random pulses with increasing levels of noise. We found that COPRA is fast, robust, and accurate. It converges reliably onto the least-squares solution for all noise levels, even from fully random initial guesses. For noisy SHG-FROG measurements we compared COPRA to PCGPA and PIE and found COPRA to be far more accurate. 

COPRA is universal and even applicable to PNPS measurements for which full amplitude and phase retrieval has not been shown before, e.g., MIIPS or SD-iFROG. Furthermore, COPRA is able to retrieve pulses from incomplete measurement traces, e.g., from iFROG traces with incomplete spectral sampling. 

We anticipate that our algorithm will have great practical value. It was designed to be easy to implement and can be directly applied to a multitude of measurements. For FROG it does not impose any relation between the frequency and delay sampling, like it is required by PCGPA. For iFROG no calculation of a subtrace is necessary as COPRA works directly with the measurement data. For d-scan it offers a reliable and fast alternative to multi-dimensional optimization. 

Some variants of COPRA remain subject of further work. For example, COPRA could be modified to work with XFROG and blind FROG. Simultaneous retrieval of the spectral response function of the measurement setup, like it was demonstrated for d-scan, could also be studied with COPRA.

Moreover, COPRA can be used as a universal and unbiased framework upon which the quality of a pulse measurement method may be judged. Which PNPS measurement is more suitable for a certain pulse can then be determined independently of the retrieval algorithm and solely based on the measurement method itself. COPRA may even be used to algorithmically engineer and optimize novel pulse retrieval methods.

In this sense, we hope that COPRA and the PNPS framework will help to give further insight in some fundamental questions of ultrashort pulse measurement: How much information is necessary for unique pulse retrieval? How large is the uncertainty in the retrieved pulse? Which PNPS method is most appropriate for certain kinds of pulses?

\paragraph{Funding} NCG acknowledges support by the German Federal Ministry of Education and Research under grant no 03ZZ0413 and grant no 03ZZ0467. TP acknowledges support by the German Research Foundation under grant no PE 1524/10-1. FE acknowledges support by the German Federal Ministry of Education and Research under grant no 13XP5053A.

\paragraph{Acknowledgement} NCG thanks JGG for providing the impetus to finally write up this manuscript.

\bigskip \noindent See the supplementary material for supporting content.

{\small
\bibliography{papers}}

\ifthenelse{\boolean{shortarticle}}{%
\clearpage
\bibliographyfullrefs{papers}
}{}

\end{document}


\maketitle
\thispagestyle{fancy}
\ifthenelse{\boolean{shortarticle}}{\abscontent}{}

\section{Additional PNPS methods}
In the main paper we showed only a selection of the published PNPS methods. Additional non-collinear PNPS methods can be found in Tab.~\ref{tab:measurement_operators}. Tab.~\ref{tab:parametrization_filters} gives parametrization filters for additional collinear PNPS methods. More can be found in the literature, although they are mostly variants of schemes shown here. In non-comprehensive numerical tests we found that COPRA can be applied to all of them.

\section{Discrete Calculations}\label{sec:discrete_calculations}
The discrete calculations are performed by approximating the integral of the continuous Fourier transform by its Riemann sum. This leads to
\begin{align}
 \senv_n &= \dft{n}{k}(\tenv_k) \equiv \sum_{n} D_{nk} \tenv_k  \quad \text{with} \quad D_{nk} = \frac{\Delta t}{2\pi} \e^{\im \wb_n t_k}  \\
 \tenv_k &= \idft{k}{n}(\senv_n) \equiv \sum_{k} D^{-1}_{kn} \senv_n   \quad \text{with} \quad D^{-1}_{kn} = \Delta \wb \e^{-\im t_k \wb_n}.
\end{align}
Here, the summations are \emph{not} discrete Fourier transforms~(DFT), and $D_{kn}$ and $D^{-1}_{km}$ are \emph{not} DFT matrices. Rather, by requiring the reciprocity relation $\Delta t \: \Delta \wb = 2\pi / N$ of the grid spacings the sums can be calculated by using a DFT \emph{and} two multiplications with appropriate phase factors that depend on the simulation grid. We have
\begin{align}
    \senv_n &= \frac{\Delta t}{2\pi} \exp(\im \wb_n t_0) \sum_{k} \bigl[\exp(\im k \wb_0 \zeta) \tenv_k \bigr] \exp(\im n k \zeta), \\
    \tenv_k &= \Delta \wb \exp(\im t_k \wb_0) \sum_{n} \bigl[\exp(-\im n t_0 \zeta) \senv_n \bigr] \exp(-\im k n \zeta), \label{eq:correct_fourier_transform}
\end{align}
with $\zeta = \Delta t \Delta \wb = 2\pi/N$. Only the last exponential in the sum belongs to the DFT and the other phase factors are required to approximate the continuous Fourier transform on the chosen grid. For the specific choice of $t_0 = - \lfloor N/2 \rfloor \Delta t$ and $\wb_0 = -\lfloor N/2\rfloor \Delta \wb$ both phase factors can be replaced by circular shifts of the input and output arrays by $\pm\lfloor N/2\rfloor$. This is what is done by the functions \texttt{fftshift} and \texttt{ifftshift} that are available in some programming languages. For any other non-trivial choice of $t_0$ and $\wb_0$ a circular shift by a different amount or the explicit expression in \eqref{eq:correct_fourier_transform} has to be used. Detailed expositions of this issue can be found in~\cite{Briggs1995,Hansen2014}.

Using the discrete approximations for the Fourier transform we can then calculate the discrete PNPS signal from $\vE$. We give two examples below. In the non-collinear case it fully depends on the method. For example, for SHG-FROG it is given by
\begin{align}
 S_{mk} = \idft{k}{n}[\exp(\im \tau_m \wb_n) \: \senv_n] \: \idft{k}{n}[\senv_n] \approx \Si_{\tau_{m}}[\senv](t_k).
\end{align}
In the collinear case we define the discretely evaluated parametrization filter
\begin{align}
 \Hf_{mn} = \Hf_{\delta_m}(\wb_n). \label{eq:discrete_pfilter}
\end{align}
With that we can calculate $S_{mk}$ for all collinear methods using SHG as
\begin{align}
 S_{mk} = \idft{k}{n}[ (\dft{n}{l}[\Hf_{ml} \senv_l])^2].
\end{align}
This shows that once the calculation of SHG-iFROG is implemented, it can be used to calculate SHG-d-scan traces by simply exchanging $\Hf_{mn}$.

\begin{table}[t]
\begin{center}
\caption{\bf Signal operator for additional non-collinear schemes.}
\begin{tabular}{cl}
\hline
Method & $\Si_\delta[\senv]$ \\
\hline
SD-FROG &  $\F^{-1}\left[\e^{\im \tau \wb} \senv \right]^2 \: \F^{-1}\left[\senv \right]^*$ \\
THG-FROG & $\F^{-1}\left[\e^{\im \tau \wb} \senv \right]^2 \: \F^{-1}\left[\senv \right]$ \\
\hline
\end{tabular}
\label{tab:measurement_operators}
\end{center}
{\small The pulse delay $\tau$ is the parameter $\delta$ in these methods.}
\end{table}
\begin{table}[t]
\begin{center}
\caption{\bf Parametrization filter for additional collinear schemes.}
\begin{tabular}{lp{1.8cm}l}
\hline
Scheme & Parameter $\delta$ & $\Hf_\delta(\wb)$  \\
\hline
chirp scan~\cite{Loriot2013} & chirp $C$ & $\exp(\im C/2 \: \wb^2)$ \\
bFROG~\cite{Forget2010} & delay $\tau$ & $[1 + \exp(\im \tau \wb)] / 2$ \\
$\pi$FROG~\cite{Hyyti2017} & delay $\tau$ & $[B(\wb+\Omega_0) + \exp(\im \tau \wb)] / 2$ \\
\hline
\end{tabular}\label{tab:parametrization_filters}
\end{center}
\end{table}

\section{Gradients}\label{sec:gradients}
COPRA requires the calculation of $\nabla_n Z_m$, which is the gradient of $Z_m$ with respect to $\vE$. As $Z_m$ is not holomorphic the expression $\nabla_n Z_m$ represents a Wirtinger derivative~\cite{Kreutz-Delgado2009}. It can be calculated by viewing $\vE$ and its complex conjugate $\vE^*$ as independent variables and evaluating
\begin{align}
\nabla_n Z_m = 2\frac{\partial Z_m}{\partial \senv_n^*}.
\end{align}
Although the calculation is straightforward we think that providing the expressions here will make the implementation of the algorithm and the reproduction of our results easier. In general $\nabla_n Z_m$ takes the following form
\begin{align}
 \nabla_n Z_m = -2 \sum_{k} \dS^*\: \frac{\partial S_{mk}}{\partial \senv_n^*}  + \dS \left[ \frac{\partial S_{mk}}{\partial \senv_n} \right]^*, \label{eq:nabla_Z}
\end{align}
where
\begin{align}
\dS = S_{mk}' - S_{mk}.
\end{align}
For methods where $S_{mk}$ does not explicitly depend on $\env_n^*$, e.g., those based on SHG and THG, the first term in \eqref{eq:nabla_Z} vanishes. In the calculations we use the orthogonality of the Fourier matrices
\begin{align}
 \sum_{k} D_{mk} D^{-1}_{kn} &= \delta_{mn}.
\end{align}
Furthermore, we use that the matrices for forward and backward transformation are related by
\begin{align}
D^{-1}_{nm} &= \frac{2\pi \Delta \wb}{\Delta t} \left[D_{mn} \right]^* \label{eq:backward_to_forward}
\end{align}

\subsection{Non-collinear Methods}
We define the shifted pulse (usually denoted $E(t_k - \tau_m)$ in the FROG literature) as 
\begin{align}
 A_{mk} &= \idft{k}{n}[ \exp(\im \tau_m \wb_n) \senv_n]. \label{eq:shifted_frog_pulse}
\end{align}

\paragraph{SHG-FROG}We have $S_{mk} = A_{mk} E_{k}$. The derivative of $S_{mk}$ is
\begin{align}
 \frac{\partial S_{mk}}{\partial \senv_n} = &D^{-1}_{kn} \exp(\im \tau_m \wb_n) E_k +  D\inv_{kn} A_{mk}.
\end{align}
Plugging this result in \eqref{eq:nabla_Z} and using \eqref{eq:backward_to_forward} we obtain
\begin{align}
  \nabla_n Z_m = &-\frac{4\pi \Delta \wb}{\Delta t} \left( \exp(-\im \tau_m \wb_n) \dft{n}{k}[\dS \: E^*_{k}] + \right. \nonumber \\
 & \left. \dft{n}{k}[\dS \: A^*_{mk}] \right). \label{eq:nabla_shg_frog}
\end{align}

\paragraph{PG-FROG}We have
\begin{align}
 S_{mk} = |A_{mk}|^2 E_k.
\end{align}
which leads to
\begin{align}
  \nabla_n Z_m = &-\frac{4\pi \Delta \wb}{\Delta t} \left(2\: \exp(-\im \tau_m \wb_n) \dft{n}{k}[ A_{mk} \: \Re\left(\dS \: E^*_{k}\right)] + \right. \nonumber \\
 & \left. \dft{n}{k}[\dS \: |A_{mk}|^2] \right).
\end{align}

\paragraph{TDP} The expressions for time-domain ptychography are very similar to FROG, requiring only the replacement of $A_{mk}$ by
\begin{align}
 A_{mk} &= \idft{k}{n}[ \tilde{B}(\wb_n) \exp(\im \tau_m \wb_n) \senv_n].
\end{align}
$B(\wb)$ describes the amplitude transmission of the bandpass filter in one correlator arm. The gradient is then given by
\begin{align}
  \nabla_n Z_m = &-\frac{4\pi \Delta \wb}{\Delta t} \left( \tilde{B}(\wb_n) \exp(-\im \tau_m \wb_n) \dft{n}{k}[\dS \: E^*_{k}] + \right. \nonumber \\
 & \left. \dft{n}{k}[\dS \: A^*_{mk}] \right). \label{eq:nabla_tdp}
\end{align}

\subsection{Collinear methods}
We define the pulse in the time domain after application of the parametrization operator
\begin{align}
 C_{mk} = \idft{k}{n}[ \Hf_{mn} \: \senv_n],
\end{align}
where $\Hf_{mn}$ is the discrete evaluation of the parametrization filter from Eq.~\ref{eq:discrete_pfilter}.

\paragraph{SHG} For methods based on SHG we have
\begin{align}
 S_{mk} = (C_{mk})^2,
\end{align}
which leads to
\begin{align}
\nabla_n Z_m = -\frac{8\pi \Delta \wb}{\Delta t} \: \Hf_{mn}^* \: \dft{n}{k}[\dS C_{mk}^*].
\end{align}

\paragraph{THG} For methods based on THG we have
\begin{align}
 S_{mk} = (C_{mk})^3,
\end{align}
which leads to
\begin{align}
\nabla_n Z_m = -\frac{12\pi \Delta \wb}{\Delta t} \: \Hf_{mn}^* \: \dft{n}{k}[\dS (C_{mk}^*)^2].
\end{align}

\paragraph{SD} For collinear methods using inline SD we have
\begin{align}
 S_{mk} = \left|C_{mk} \right|^2 C_{mk},
\end{align}
which leads to
\begin{align}
\nabla_n Z_m = -\frac{4\pi \Delta \wb}{\Delta t} \Hf_{mn}^* \dft{n}{k}\left[ \dS^* \: C_{mk}^2 + \right. \nonumber \\ 
\left. 2 \: \dS \: |C_{mk}|^2 \right].
\end{align}

\subsection{Global iteration}
The expression for the gradient $\nabla_{mk} r$ used in the global iteration was obtained using exactly the same method as above. The global gradient $\nabla_n Z$ is obtained by simply summing the above expressions over $m$
\begin{align}
 \nabla_n Z = \sum_{m} \nabla_n Z_m.
\end{align}

\subsection{Implementation}
Nowadays, most programming languages offer a data type for complex numbers which can be used to implement the expressions above straightforwardly. This is what makes the Wirtinger formalism so attractive. The alternative of giving the expressions in derivatives of $\Re[\senv_n]$ and $\Im[\senv_n]$ would make the notation and implementation more cumbersome.

Because $\dS$ has to be calculated in the algorithm before $\nabla_k Z_m$, the values $\dS$, $A_{mk}$, $E_n$, $C_{mk}$ and $\Hf_{mn}$ can be reused. As a result the calculation of the gradient is relatively cheap in terms of computing time. It requires only one (collinear methods) or two (non-collinear methods) additional fast Fourier transforms~(FFT).

The scaling factors in front of the gradients are inconsequential and specific to the Fourier transform convention that was used. We provide them merely for completeness.

\section{Methods}
\subsection{Test pulses}\label{sec:test_pulses}
\begin{figure}[tb]
    \centering
    \includegraphics{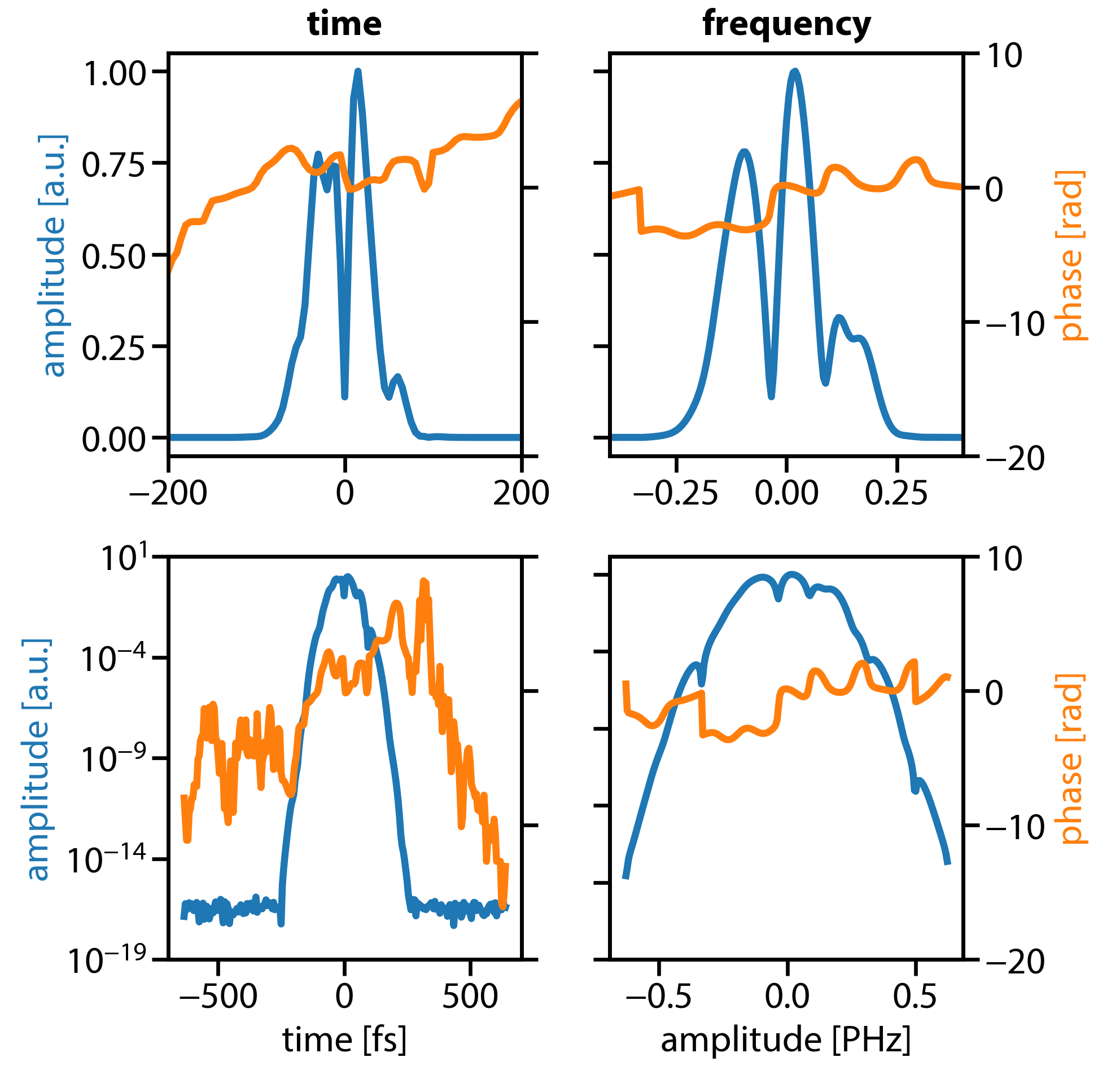}
    \caption{An example for one of the test pulses with TBP 2. In the bottom row we show the logarithmic scaling to demonstrate the time-frequency localization.}
    \label{fig:test_pulses}
\end{figure}
We aimed to create test pulses that are localized in time and frequency while having a complex amplitude and phase structure in both domains. In this way they serve as a good benchmark on how algorithms pick up spectral \emph{and} temporal amplitude features. One example is shown in Fig.~\ref{fig:test_pulses}.

We found the time-frequency localization to be especially important. If the pulse extends to the edges of the simulation grid in any domain, wrap-around will occur in the computation of the trace and the result will usually be unphysical. This problem is made worse when computing d-scan and MIIPS traces as there the pulse is spectrally chirped and, consequently, stretched in the time domain. Curiously, this can improve the conditioning of the retrieval problem and can lead to faster convergence of COPRA -- however, the resulting solution is unphysical~(see Sec.~\ref{sec:random_initial_guesses}). 

To avoid any issues like that we created pulses that dropped off to a relative amplitude of $\num{1e-15}$ at the edges of the simulation grid in both domains. Additionally, we had to leave enough padding in the time domain so that even the strongly chirped pulses in d-scan could be calculated without wrap-around. This very strict requirement explains why we chose a comparably low TBP of $2$ for our simulations. For higher TBPs we would have needed much larger grid sizes, e.g., $N=1024$ for a TBP of $5$. This would have made the simulation of the retrieval of a very large number of pulses infeasible at some point -- even when using COPRA. By limiting the TBP we can make sure that the computed traces do not suffer from numerical artifacts and the reported retrieval probabilities are meaningful for the retrieval from experimental traces. An example for retrieval of a pulse with very high TBP is shown in Sec.~\ref{sec:high_tbp}.

The procedure to create the test pulses requires a target TBP and a simulation grid. First, a random array is created in the frequency domain~(amplitude uniformly distributed on $[0, 1]$, phase uniformly distributed on $[0, 2\pi]$). This array is multiplied by a Gaussian function that just reaches the required edge values at $\wb_0$ and $\wb_{N-1}$. Then the Fourier transform is calculated and the result is multiplied by a Gaussian function in the time domain. The end result is the test pulse in the time domain.

The width of the second, temporal Gaussian function is roughly determined by the Fourier transform of the spectral one. However, its exact width has to be optimized to result in the specified TBP exactly. This is done with a scalar minimization algorithm.

\subsection{General minimization algorithms}
As a complementary approach to COPRA we suggested using general minimization algorithms. This was illustrated with a retrieval comparison from a simple FROG measurement using different minimization algorithms.

Specifically, we used the implementations from the SciPy package for Python in its version 1.1.0~\cite{Jones2001--}. The NM and BFGS algorithms were run from the scalar minimization wrapper \texttt{minimize} with default arguments. We used the DE implementation in the function \texttt{differential\_evolution} and the parameters described in~\cite{Escoto2018}. We used the sophisticated LM implementation that is provided as the default solver of the \texttt{least\_squares} function with its default arguments. We found that other implementations of LM may show slightly worse convergence speed~(e.g., the one in MINPACK).

The LM algorithm can be provided with an implementation of the Jacobian of $T_{mn}$ for more accurate and efficient retrieval. The required expressions can be obtained using the methods from Sec.~\ref{sec:gradients}. However, we found that in this case there is no great advantage in doing so. The calculation of the full Jacobian is computationally expensive and the numerical finite-differences approximations worked very well for us.

On the other hand, the gradient of $r$ required for BFGS can be calculated efficiently -- either by symbolic or automatic differentiation. The resulting speed-up makes BFGS an attractive alternative to LM if the extra effort of implementing the gradient is invested. This approach was not investigated in this work.

\subsection{COPRA retrieval tests}
\begin{table}[tb]
\begin{center}
\caption{\bf Parameters for our retrieval simulations.}
\begin{tabular}{p{2cm}p{4cm}}
\hline
Methods & Parameters \\
\hline
PG-FROG SHG-FROG SHG-iFROG THG-iFROG & $\tau_m = t_m$\newline  $M=N$ \\
\hline
SD-iFROG & $\tau_m = t_0 + m \Delta \tau$ \newline $\Delta t = (t_{N-1} - t_{0})/M$\newline $M=4N$ \\
\hline
SHG-TDP & $\tau_m = t_0 + m \Delta \tau$\newline $\Delta t = (t_{N-1} - t_{0})/M$\newline $M=128$ \\
\hline
SHG-d-scan THG-d-scan SD-d-scan & $z_m = (m - M/2 + 0.5) \Delta z$\newline $\Delta z = \SI{25.0}{\mm}/M$\newline $M=128$ \newline glass: BK7\\
\hline
SHG-MIIPS THG-MIIPS SD-MIIPS & $\delta = m \Delta \delta$\newline $\Delta \delta = 2\pi /M$\newline $M=128$\newline $\gamma = \SI{22.5}{\fs}$\newline $\alpha=1.5 \pi$ \\
\hline
\end{tabular}
\label{tab:parameters}
\end{center}
\end{table}
The parameters we used for our pulse retrieval simulations for the different methods are listed in Tab.~\ref{tab:parameters}. The parameters were chosen after careful consideration of several factors. For SHG-FROG we wanted to do a direct comparison to PCGPA which requires the delay sampling $\tau_m = t_m$. The same was also chosen for PG-FROG, SHG-iFROG and THG-iFROG. For SD-iFROG we found that higher delay sampling is necessary for reliable convergence. For SHG-TDP we used only $M=128$ delay samples.

For d-scan the maximum glass insertion and the pre-chirp were adapted to the TBP of the pulse, because retrieval worked best when the largest induced GVD was on the order of the pulse chirp. The pre-chirp was simulated by using negative insertion distances. This leads to a slight difference from a true d-scan measurement where pre-chirping is usually done in a grating compressor. However, we found this difference to be inconsequential from the perspective of the retrieval algorithm. For MIIPS we also adapted the parameters $\alpha$ and $\gamma$ to match the spectral chirp of the pulse.

In general, the parameters we selected are biased in the sense that retrieval for all methods should be possible within the iteration count chosen for the simulation. It is certainly possible to choose parameters that are  harder to retrieve. For example, for SD-iFROG for $\tau_m = t_m$ over $1000$ iterations would have been necessary to reach comparable trace errors. However, these cases touch the question of which minimal information is necessary for pulse retrieval -- something that we did not look at in this work.

While we tried to use parameters that are similar to the experiment this was clearly not possible for all PNPS schemes. Our actual values for bandwidth and center wavelength were taken from pulses in our lab and do not correspond to single-cycle pulses. Thus, e.g., the required glass insertion for d-scan is far higher than what has been used in an experiment (and probably also higher than what is feasible). But we stress again that the retrieval is agnostic to these specific numbers as the whole trace and the simulation can be scaled and shifted to different frequencies. Our results should be directly transferrable to the retrieval of single-cycle pulse measurements.

\subsection{Ambiguity removal}
After retrieval from synthetic PNPS traces we want to compare the retrieved pulse to the original that was used to create the trace. We found it surprisingly hard to develop a reliable procedure for this task which is why we present it here. We hope that others may find it useful for their own work. Similar methods for ambiguity removal are described in~\cite{Sidorenko2016,Maiden2017}.

In the following we will adopt a notation in which arrays are denoted by bold font~(e.g., $\vE$) and all operations are defined element-wise~(e.g., $\mu \vE$ and also $\vE \vE^0$). We assume that all arrays have $N$ elements and define $\vec{\wb} = (\wb_0, \ldots, \wb_{N-1})$. This notation is imprecise but it should allow an easy implementation in modern programming languages. We start by defining the NRMSE as a function of two arrays
\begin{align}
 \nrmse(\vec{x}, \vec{y}) = \bigl[ \sum_{n=0}^{N-1} |x_n - y_n|^2 / ( N \max_{n} |y_n|^2 )\bigr]^{1/2}.
\end{align}
The retrieval error $\varepsilon$ can then be defined as
\begin{align}
    \varepsilon(\vE) \equiv \min_{\mu, \varphi_0, \varphi_1} \nrmse\{\mu \exp[\im (\varphi_0 + \varphi_1 \vec{\wb})] \vE, \vE^0\}.
\end{align}
To calculate it we need two analytical results. First the error
\begin{align}
 &\nrmse(\mu |\vE|, |\vE^0|) \\
\shortintertext{is minimized by}
\mu(\vE, \vE^0) &= \bigl( \sum_{n} |\senv_n| \: |\senv_n^0| \bigr) / \bigl( \sum_n |\senv_n^0|^2 \bigr).
\end{align}
This calculates the best scaling to match the amplitudes of two complex-valued arrays in the least-squares sense. It can be obtained by straightforward differentiation. Additionally, the error
\begin{align}
 &\nrmse(c_0 \vE, \vE^0) \quad \text{with} \quad |c_0| = 1, c_0 = \exp(\im \phi_0) \\
\shortintertext{is minimized by}
c_0'(\vE, \vE^0) &= \bigl( \sum_{n} \senv_n^0 \senv_n^* \bigr) / \bigl| \sum_{n} \senv_n^0 \senv_n^* \bigr|, \\
c_0(\vE, \vE^0) &= \begin{cases}
    c_0'(\vE, \vE^0) & \nrmse(c_0 \vE, \vE^0) < \nrmse(-c_0 \vE, \vE^0) \\ 
    -c_0'(\vE, \vE^0) & \text{else}
\end{cases}
\end{align}
This calculates the optimal constant phase to match two complex-valued arrays in the least-squares sense. Using these results we can proceed to calculate $\varepsilon$. First $\vE$ is scaled to match the amplitude of $\vE^0$:
\begin{align}
  \vE \to \mu(\vE, \vE^0) \vE.
\end{align}
Then we define the objective function
\begin{align}
 O(\phi_1) &= \nrmse\{ c_0(\vE',\vE^0) \: \vE',\:\: \vE^0\} \\
\shortintertext{with}
\vE' &\equiv \vE'(\phi_1) = \exp(\im \phi_1 \vec{\wb}) \vE.
\end{align}
This calculates the minimal retrieval error $\epsilon$ for a given linear spectral phase $\phi_1$. It remains to determine the optimal $\phi_1$. For that we sample $O(\phi_1)$ using $2N$ regularly spaced points between $-\pi / \Delta \wb$ and $\pi / \Delta \wb$ to obtain a bracket that encloses the global minimum. The exact location of the minimum is found by a bisection method. The final retrieval error is $O$ at the optimal $\phi_1$. The direction-of-time ambiguity is taken into account by trying out both variants and selecting the one with the lower retrieval error, i.e., $\epsilon = \min[\epsilon(\vE), \epsilon(\vE^*)]$.

\section{Common pulse retrieval algorithm}\label{sec:copra}
\begin{figure}[tb]
    \centering
    \includegraphics{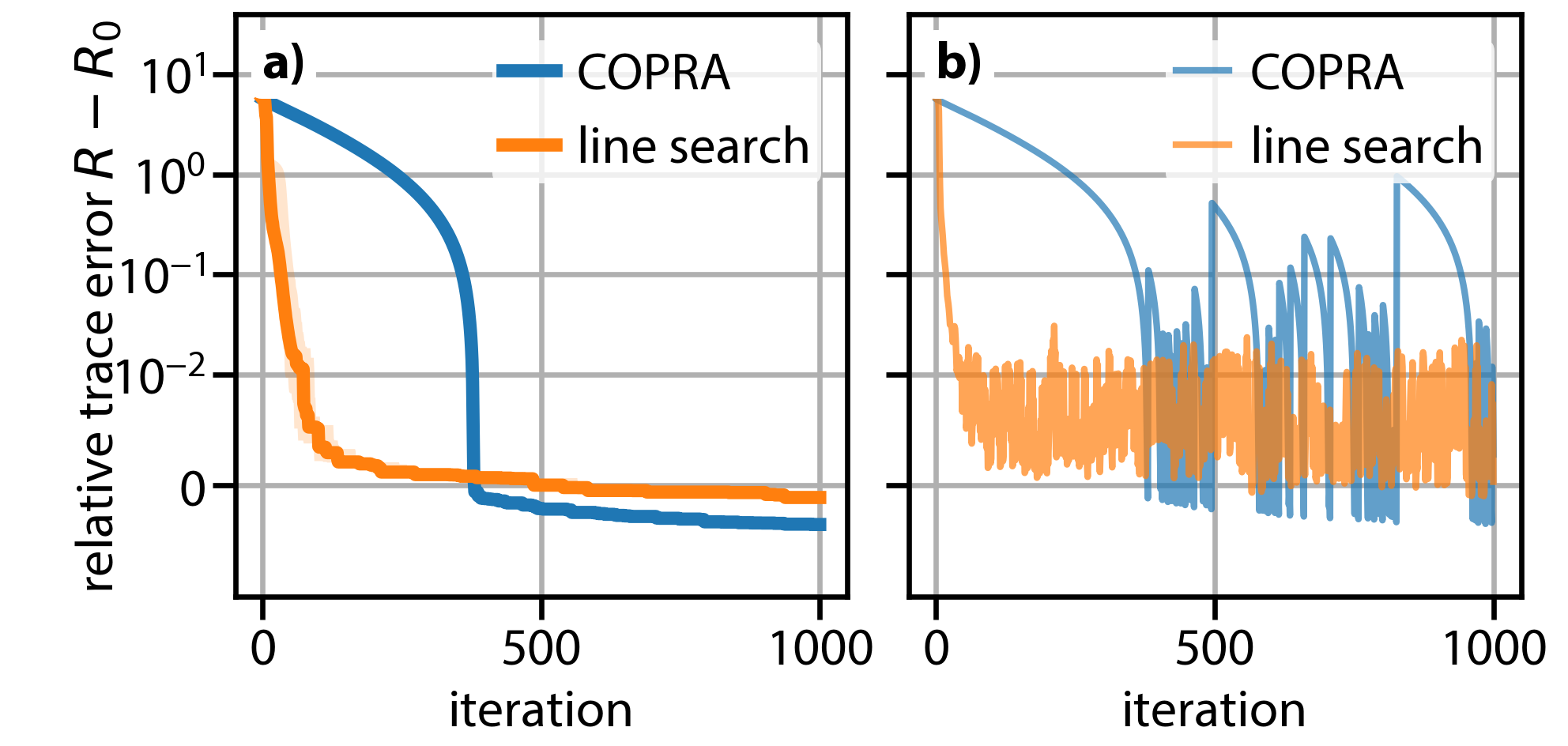}
    \caption{Convergence behavior for a noisy SHG-FROG trace using the global iteration of COPRA and a modified version using a backtracking line search. a) the running minimum of the trace error b) the actual trace error over the iteration count.}
    \label{fig:backtracking_comparison}
\end{figure}
In this section we want to discuss various aspects of COPRA in more detail. This includes design decisions and possible modifications.

\paragraph{Trace error}
One aspect that was silently disregarded in the main paper is that the trace error $R$ cannot be calculated during the local iteration without extra effort. As $\vE$ is changed after processing every spectrum, $S_{mn}$ is calculated based on a different $\vE$ for every $m$. Still, we recommend calculating an approximation to $R$ by using the $S_{mn}$ from the local iteration. As the changes in $\vE$ between the iterations become smaller it will get closer to the real $R$. We observed that it works very well to use this approximation to terminate the local iteration and to select the best solution. All results from the main paper were all obtained in this way. 

However, one has to be very careful when reporting the error $R$ during or after the local iteration of COPRA. The approximation will regularly underestimate $R$. Similarly, one has to take care to calculate the correct final trace. It cannot be based on the $S_{mn}$ obtained during the local iteration. Rather, it has to be calculated in full for the retrieved pulse $\vE$. In this work, we went to the extra effort and calculated the correct $R$ in every local iteration for logging purposes~(this data is shown in the convergence graphs).  

We remark that these problems also exist in ptychographical algorithms.
\paragraph{Convergence behavior}
The convergence graphs in the main paper explicitly show the \emph{running minimum} of the trace error $R$. Ignoring that, one may get the wrong impression that COPRA reduces the trace error in every iteration. This is not the case. Rather the trace error can fluctuate strongly between iterations. Still, the minimal trace error $R$ will converge. An example of the actual behavior of the trace error in dependence of the iteration count is shown in Fig.~\ref{fig:backtracking_comparison}~b).

\paragraph{Projection}
The projection on the measured intensities is denoted in the following way in the main paper:
\begin{align}
    S_{mk}' = \mu \: \idft{k}{n} \left( \tilde{S}_{mn} / | \tilde{S}_{mn} |  \: \sqrt{T^\mathrm{meas}_{mn}}\right), \label{eq:projection}
\end{align}
Two modifications have to be made for implementation. First, to avoid a division by zero we test for the magnitude of $|\tilde{S}_{mn}|$ beforehand. Elements that are smaller than $N\epsilon_\mathrm{m} \max|\tilde{S}_{mn}|$ where $\epsilon_\mathrm{m}$ is the machine accuracy are set to $1$. Second, the complex-valued square root is calculated, effectively mapping negative intensities to imaginary values. These negative values appear naturally when measuring with CCD spectrometers and subtracting the dark count. It was already observed in the literature that this way of handling them is necessary for an unbiased projection, e.g., in ~\cite{Miranda2017}.

\paragraph{Spectra iteration order} We recommended to iterate randomly over the spectra in the local iteration stage without commenting on the effects. We found that it generally has a positive effect on the retrieval ratio. However, depending on the PNPS scheme other iteration orders can be more beneficial, e.g., iterating the spectra from highest to lowest intensity.  However, the effect is small and any iteration order should lead to comparable results.

\paragraph{Step sizes}
The speed of COPRA benefits from not requiring a line search in every iteration. Rather, we give heuristically safe step sizes $\gamma$, $\eta_z$, and $\eta_r$ that show good convergence in our tests. Our step sizes are all based on one specific choice, namely the function value divided by the L2-norm of its gradient. For example, in the global iteration we have 
\begin{align}
    \senv_n' = \senv_n - \alpha \bigl( Z / \sum_k | \nabla_k Z |^2 \bigr) \nabla_n Z.
\end{align}
We motivated this step size by performing a Taylor expansion. For example, we expand $Z(\vE')$ in $\vE$ using $\eta$ as the step size~\cite{Kreutz-Delgado2009}:
\begin{align}
 Z(\vE') &\approx Z(\vE) + 2 \Re\bigl[ \sum_{n} (\senv_n' - \senv_n)^* \: \nabla_n Z \bigr] \\
 Z(\vE') &\approx Z(\vE) + 2 \Re\bigl[ \sum_{n} - \gamma (\nabla_n Z)^* \: \nabla_n Z \bigr] \\
 \Rightarrow \quad \eta &\approx \bigl(Z(\vE) - Z(\vE') \bigr) / \bigl(2 \sum_{n} | \nabla_n Z |^2 \bigr). \\
 \intertext{Now we assume that every iteration substantially decreases $Z$, i.e., $Z(\vE') \ll Z(\vE)$ to obtain}
 \eta &\approx Z(\vE) / \bigl(\sum_{n} | \nabla_n Z |^2 \bigr),
\end{align}
which is the choice we used in COPRA, up to a constant scale. Though we have not found any explicit discussion of this choice in the literature, it is similar to what is recommended as the starting value for a line search in some algorithms~\cite{Wright1999}.

For application in COPRA we scale this $\eta$ by a factor $\alpha$. In the local iteration we use $\alpha = 1$ (and name it $\gamma$ to allow subscripts), in the global iteration by $\alpha = 1/4$ (to obtain $\eta_z$ and $\eta_r$). We found that for many PNPS methods $\alpha$ can be safely increased for the local iteration. This increases the convergence speed but may lead to divergence in some cases. For the global iteration $\alpha$ can be decreased to obtain a more accurate solution. However, this obviously slows down the convergence.

We tested more sophisticated approaches where $\nu$ is decreased over the iteration count, e.g, $\alpha \propto (1.0 - \mathrm{iter.}/\text{max. iter.})$. Sometimes this leads to faster and more accurate convergence, but we did not find a generally applicable strategy.

We also compared our step sizes to those obtained by an inexact line search. Specifically, we modified the global iteration to use a backtracking line search that fulfills the Armijo condition~(see \cite{Wright1999} p. 37) with $c=\num{1e-4}$ and $\rho = 0.75$. It was used for both the minimization of $r$ and $Z$. Then we ran only the global iteration of COPRA on a noisy SHG-FROG trace. The results are shown in Fig.~\ref{fig:backtracking_comparison}. We see that the line search initially converges faster -- in terms of iteration count, not function evaluations -- but then stagnates above the accuracy reached by the original COPRA. However, in Fig.~\ref{fig:backtracking_comparison}~b) we see that the trace error in COPRA fluctuates more strongly. An exact line search could provide additional accuracy but we conclude that the COPRA step sizes are generally an almost optimal choice for the global iteration.

For the local iteration we observed the same, almost optimal behavior only in the noiseless case. When noise is present adapting the stepsize to the local gradient norm results in poor convergence and even leads to divergence. The modification we provided, the usage of the maximum gradient norm in the denominator instead of the local gradient norm, is purely heuristic. We cannot provide any more insight except than that we observed it to work very well.

\paragraph{Ambiguities in pulse retrieval}
It seems that even though the pulse retrieval problem is ill-posed due to its many ambiguities solving it does not require to remove the ambiguities in retrieval. We observed no clear indication of stagnation or a cyclic behavior due to ambiguous solutions like it happens in classical phase retrieval. In practice, COPRA will simply converge randomly to one of the many ambiguous solutions. On the other hand, our attempts to remove the ambiguities, e.g.,  by fitting the second derivative of the phase, showed no improvement.

In fact, COPRA even seems to profit slightly when the ambiguities are used to effectively widen the search range. After every iteration we can update the current guess by
\begin{align}
 \senv_n' = \exp[\im(\varphi_0 + \varphi_1 \wb_n)] \senv_n,
\end{align}
where $\phi_0$ and $\phi_1$ are randomly chosen. We found that this can slightly increase the retrieval probability -- but the increase was not significant enough to be included in the default algorithm.

This effect was used when we introduced an additional ambiguity to the algorithm by the scale factor $\mu$ in the expression for $r$. It leads to an effective ambiguity in the scale of $\vE$. Originally, it should only serve to assure convergence even when an initial guess with the wrong magnitude was provided (by determining $\mu$ once in the first iteration). But we observed that convergence is more reliable when it is updated in every iteration.

\paragraph{Momentum} The convergence speed of COPRA can be increased further by calculating an exponentially-decaying sum of the past gradients and using that as the descent step. This is usually likened to a momentum of the iteration and is well-known to increase the learning rate in neural networks~\cite{Ruder2016}. This approach is also used in the extended ptychographical iterative engine~\cite{Maiden2017}. However, as COPRA often converges within 100~iterations we found this additional increase in speed not worth the effort. Thus, it was excluded from the version presented in the paper for brevity. Still, it may be useful for very large retrieval problems where a single iteration is already computationally expensive.

\paragraph{Weighted least-squares}\label{subsec:weighted_copra}
In the main paper we studied the influence of constant additive Gaussian measurement noise on the pulse retrieval. This case is commonly assumed in pulse retrieval simulations~\cite{Sidorenko2016,Hyyti2017} and corresponds to low intensity measurements with CCD array spectrometers as they are often used for PNPS measurements~\cite{Davenport2015}. In other situations the signal-dependent part of the noise in a measurement will dominate.

COPRA can be modified to solve the weighted least-squares problem to deal with this case. It requires knowledge of the standard deviation $\sigma_{mn}$ of the Gaussian noise of the measurement $T_{mn}$, e.g., from repeated measurements. In this case a maximum-likelihood estimate can be obtained by solving the weighted least-squares problem:
\begin{align}
 r = \sum_{m,n} \left[\Tm_{mn} - \mu T_{mn}(\vE) \right]^2 / \sigma_{mn}^2, \label{eq:residuals}
\end{align}
Making COPRA work with this modified $r$ two small changes. First, the expression for $\mu$ has to be modified to
\begin{align}
    \mu = \left[\sum_{mn}  (\Tm_{mn} \: T_{mn}) / \sigma_{mn}^2 \right] / \left[ \sum_{mn} T_{mn}^2 / \sigma_{mn}^2 \right]. \label{eq:scaling_factor}
\end{align}
Second, in the global iteration the expression for the gradient of $r$ has to be adapted to include the weighting
\begin{align}
    \nabla_{mk} \nnR  = -4 \mu \frac{\Delta t}{2\pi \Delta \omega} \idft{k}{n} \left[ \bigl( \Tm_{mn} - \mu T_{mn} \bigr) \tilde{S}_{mn} / \sigma_{mn}^2 \right].
\end{align}
After these modifications COPRA will solve the weighted least-squares problem.

Analyzing the retrieval in presence of more complicated noise models, e.g., the combination of additive and multiplicative noise or if the $\sigma_{mn}$ are not known, is out of the scope of this work. We can only say that if the $\sigma_{mn}$ are known or approximated well enough the solution obtained by COPRA will again be a maximum-likelihood estimate -- and more accurate than the results obtained by other algorithms. We also point out that the impact of moderate levels of multiplicative Gaussian noise on PCGPA or PIE is much smaller than that of additive Gaussian noise. Consequently, the advantage of using COPRA in these situations is not as large.

\section{Pulse retrieval algorithms}
Here we describe other fast pulse retrieval algorithms, some of which were compared against COPRA. It mainly serves to provide information on how we implemented them and to discuss their properties. Furthermore, we present them using the notation of the PNPS formalism.

All the algorithms in this section work by projecting on the measured intensity $\Tm$, i.e., $S_{mk}'$ is obtained by \eqref{eq:projection}
either for all $m$ in parallel or subsequently for one $m$ at a time. The scale factor $\mu$ was included by us as we found it improves convergence for all these algorithms. Its calculation is described in the main paper. For the treatment of zero and negative intensities we refer to the discussion in Sec.~\ref{sec:copra}.

\subsection{Generalized projections algorithm}
The generalized projections algorithm~(GPA) for FROG~\cite{DeLong1994} can be applied to all FROG variants. We will shortly sketch the SHG version here. 

Every GPA iteration starts with calculating $S_{mk}$ from the current solution $\vEt$ in the time domain. Then a better guess for the PNPS signal $S_{mk}'$ is obtained by projecting on the measurement via \eqref{eq:projection} (data constraint). In the next step $\vEt$ is updated by minimizing 
\begin{align}
 Z = \sum_{mk} |S_{mk}' - S_{mk}|^2
\end{align}
in terms of $\vEt$ by a gradient descent step
\begin{align}
 \tenv_j' = \tenv_j - \gamma \nabla_j^t Z. \quad \text{mathematical-form constraint}
\end{align}
Until now no assumption has been made about the delays $\tau_m$ or their sampling frequency $\Delta \tau$. An in fact, none is required. The general expression for $\nabla_j^t Z$ can be calculated with the methods described in Sec.~\ref{sec:gradients} and is given by
\begin{align}
 \nabla_j^t Z &= \sum_{m} 2\frac{\partial Z_m}{\partial \tenv_j^*} \\
    &= -2 \sum_{m} \bigl\{ \Delta S_{mj} A_{mj}^* + \nonumber \\
    &\qquad\quad \sum_{k} \Delta S_{mk} \bigl[\tenv_k \sum_n D_{kn}^{-1} D_{nj} \exp(\im \tau_m \wb_n) \bigr]^* \bigr\}, \label{eq:full_frog_gradient}
\end{align}
with the definitions from Sec.~\ref{sec:gradients}. We see that this general gradient expression is complicated and direct evaluation in this form would require a multiplication with a dense $N\times N$ matrix in the second term. To avoid this issue the gradient in COPRA is calculated with respect to $\vE$. In GPA the solution is to require $\tau_m = t_m$, which was later misunderstood as a fundamental requirement for FROG measurements. With this specific choice of delays we can simplify the gradient to obtain
\begin{align}
 \nabla_j^t Z &= -2 \sum_{m} \Delta S_{mj} A_{mj}^* + \Delta S_{mj'} E_{j'}^* \label{eq:gradient_gpa} \\
 &\quad \text{with}\quad j' \equiv j'(k, m) = \begin{cases} j + m & j + m < N \\ j + m - N & j + m \geq N \end{cases},
\end{align}
where the second term is circularly shifted by $-m$. Furthermore, $A_{mj}$ can be expressed as a circular shift of $E_k$ by $m$:
\begin{align}
 A_{mj} = E_{j'} \quad \text{with}\quad j' \equiv j'(k, m) = \begin{cases} j - m & j \geq m \\ N + j - m & j < m  \end{cases}. \label{eq:cyclic_shift}
\end{align}
Using this and assuming a periodic continuation of the involved fields we can express the gradient as 
\begin{align}
 \nabla_j^t Z &= -2 \sum_{m} S'(t_j, \tau_m) E^*(t_j - \tau_m) - E(t_j) |E(t_j -\tau_m)|^2 + \nonumber\\
 &\quad S'(t_j + \tau_m, \tau_m) E^*(t_j + \tau_m) - E(t_j) |E(t_j + \tau_m)|^2, \label{eq:gradient_gpa_continuous}
\end{align}
which is a notation more common in the FROG literature~(compare \cite{Trebino2002}, pp.\ 172--173). The step size $\gamma$ in classical GPA is determined by an exact line search. We note that a more efficient version of GPA can be implemented by using similar step sizes as the ones in COPRA, described in Sec.~\ref{sec:copra}. We did not include GPA in our comparison as we found the algorithm described in the next section to provide a better overall performance.

\subsection{Principal components GPA}
The principal components generalized projections algorithm~(PCGPA)~\cite{Kane1998} is inspired by GPA and also requires the specific delay sampling $\tau_m = t_m$. After obtaining $S_{mk}'$ by \eqref{eq:projection} it uses the algebraic structure of the FROG trace to update $\vEt$. Specifically, we can obtain the so-called outer product form from $S_{mk}$ by reversing and performing a circular shift by $k$ for every $k$th column:
\begin{align}
  S_{mk} &\to \hat{S}_{mk} \quad \text{under} \: m \to \begin{cases} k - m & k \geq m \\ N + k - m & k < m \end{cases} \\
  \intertext{which is}
  \hat{S}_{mk} &= E_{m} E_{k}.
\end{align}
This is a direct result from \eqref{eq:cyclic_shift}. For the corrected PNPS signal $S_{mk}'$ this outer product form will not exactly match to a field $\vEt$, however, it can be decomposed in a least-squares sense by a singular value decomposition~(SVD). As this approach scales badly with the size of the trace, PCGPA commonly uses a single iteration of the power method to obtain an estimate for the eigenvector associated with the largest eigenvalue. Specifically, this means to perform the following steps to update $\vEt$
\begin{align}
 \tenv_k &\to \sum_k (\hat{S}_{mk}')^* \tenv_k \\
 \tenv_k &\to \tenv_k^* / \sum_l |\tenv_l |^2.
\end{align}
PCGPA can be modified to work with other FROG variants in which case the formulas above would have to be adapted. We found that using a full SVD gives more accuracy and slightly improved retrieval probability. So for small grid sizes it may the preferable strategy.

\subsection{Ptychographic iterative engine}
Recently, several pulse retrieval algorithms based on ptychography have been proposed~\cite{Spangenberg2015,Witting2016,Sidorenko2016,Sidorenko2017}. All of them are based on the ptychographic iterative engine~(PIE)~\cite{Maiden2009,Maiden2017}, which we chose as a common name for these slightly different algorithms. Specifically, in our comparisons we used the version for SHG-FROG from~\cite{Sidorenko2016,Sidorenko2017} which we describe in the following.

In PIE for SHG-FROG the spectra are processed individually and in random order. For every $m$ the PNPS signal $S_{mk}$ is calculated from the current guess $\vE$, followed by a projection on the measured data to obtain $S_{mk}'$. Then an updated guess for $\vEt$ is obtained by
\begin{align}
 \tenv_k' = \tenv_k + \beta \bigl( A_{mk}^* / \| \vEt \|_\mathrm{max}^2 \bigr) \dS, \label{eq:ptychography} 
\end{align}
where $\beta \in [0.1, 0.5]$. Comparing this expression to \eqref{eq:full_frog_gradient} we see that it is, in fact, a gradient descent step using only the first term of the GPA gradient with the specific step size
\begin{align}
 \gamma = \beta / \| \vEt \|_\mathrm{max}^2.
\end{align}
The relation of the update step in ptychography to gradient descent has been discussed in~\cite{Maiden2017}. There this specific step size is identified as the Lipschitz constant of the gradient.

The fundamental reason why only one part of the GPA gradient is used, is that in ptychography probe and object pulses, i.e., the delayed and undelayed pulses, are seen as independent variables. Now in PIE for XFROG and TDP both pulses are updated separately using the gradient step from above, which effectively means using both terms of the full gradient. This is very similar to GPA for blind FROG~\cite{Trebino2002}.

However, in PIE for SHG-FROG only a single gradient step is performed. For the specific choice $\tau_m = t_m$ this has almost no impact as then both parts of the GPA gradient are approximately the same under the transformation $\tau_m \to -\tau_m$. This can be seen best from \eqref{eq:gradient_gpa_continuous}. Therefore, if $\tau_m$ and $-\tau_m$ are processed subsequently the algorithm still picks up the full gradient approximately. However, when the delay sampling is not symmetrical, e.g., $\tau_m = t_m + \Delta t/2$ or even $\tau > 0$, we found that the retrieval probability is reduced significantly.

The discussion here should make it clear that from a numerical perspective there is no large conceptual difference between GPA and PIE. The only major distinction is that PIE processes the spectra individually and provides a specific step size for the gradient. In general, the results obtained by both algorithms are also comparable -- something that was confirmed in our work.

The slightly improved retrieval accuracy of PIE compared to GPA in some cases~\cite{Sidorenko2017} stems from the individual processing of the spectra and, specifically, the order in which they are iterated. While in GPA all spectra contribute equally, in PIE the spectra that are processed last effectively influence the solution more strongly. So depending on the processing order the retrieval accuracy can be slightly better or slightly worse.

\subsection{GPA for d-scan}
Recently, a fast retrieval algorithm for d-scan based on GPA has been proposed~\cite{Miranda2017}. It proceeds by projecting on the measured intensity and then updates $\vE$ by a heuristic procedure:
\begin{align}
 S_{mk}' &\to S_{mk}' \: \idft{k}{n}[\Hf_{mn}^* \senv_n^*] \\
 S_{mk}' &\to S_{mk}' / |S_{mk}'|^{2/3} \\
 \senv_n &\to \sum_{m} \Hf_{mn}^* \dft{n}{k}[S_{mk}']
\end{align}
We found that when using it for simultaneous amplitude and phase retrieval this algorithm does not converge by the definition used in this work. For noiseless SHG-d-scan traces we found it never obtains $R < \num{5e-3}$ or $\varepsilon < 1\%$. If seeded with the original test pulse it quickly diverges from the numerical limit of $R \approx \num{1e-15}$ and settles at $R \approx \num{5e-3}$. This means that the solution of the pulse retrieval problem is not a fixed point of the algorithm. For these reasons it was not included in a comparison.

\section{Random initial guess}\label{sec:random_initial_guesses}
\begin{figure}[tb]
    \centering
    \includegraphics{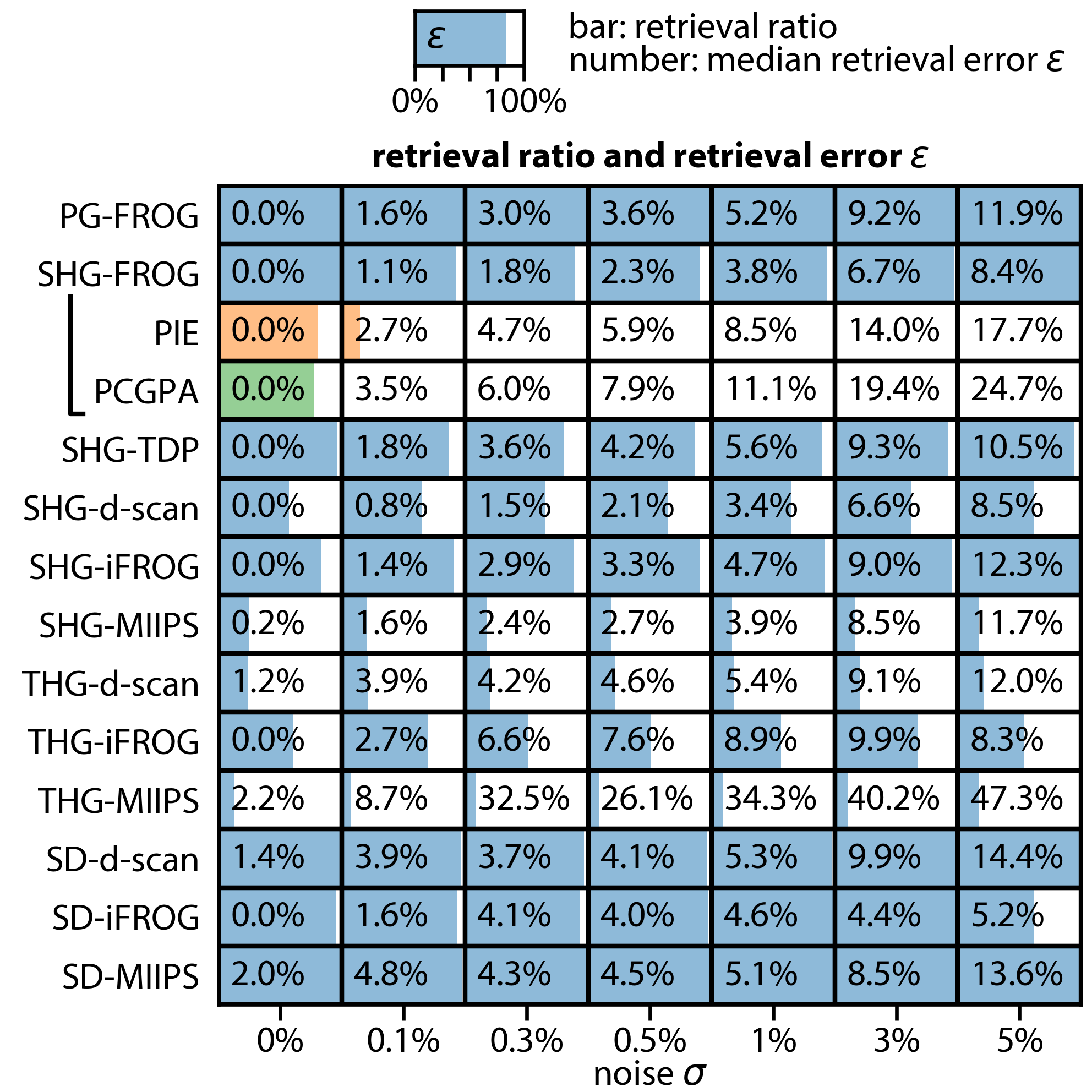}
    \caption{The retrieval ratio (blue bar) and median retrieval error (inset number) of COPRA in dependence of the PNPS method and the noise level \emph{when starting with a random initial guess}. A comparison with PCGPA and PIE for SHG-FROG is included. Successful retrieval was assumed if $R < R_0 + \num{1e-4}$.}
    \label{fig:random_retrieval_probabilities}
\end{figure}
\begin{figure}[tb]
    \centering
    \includegraphics{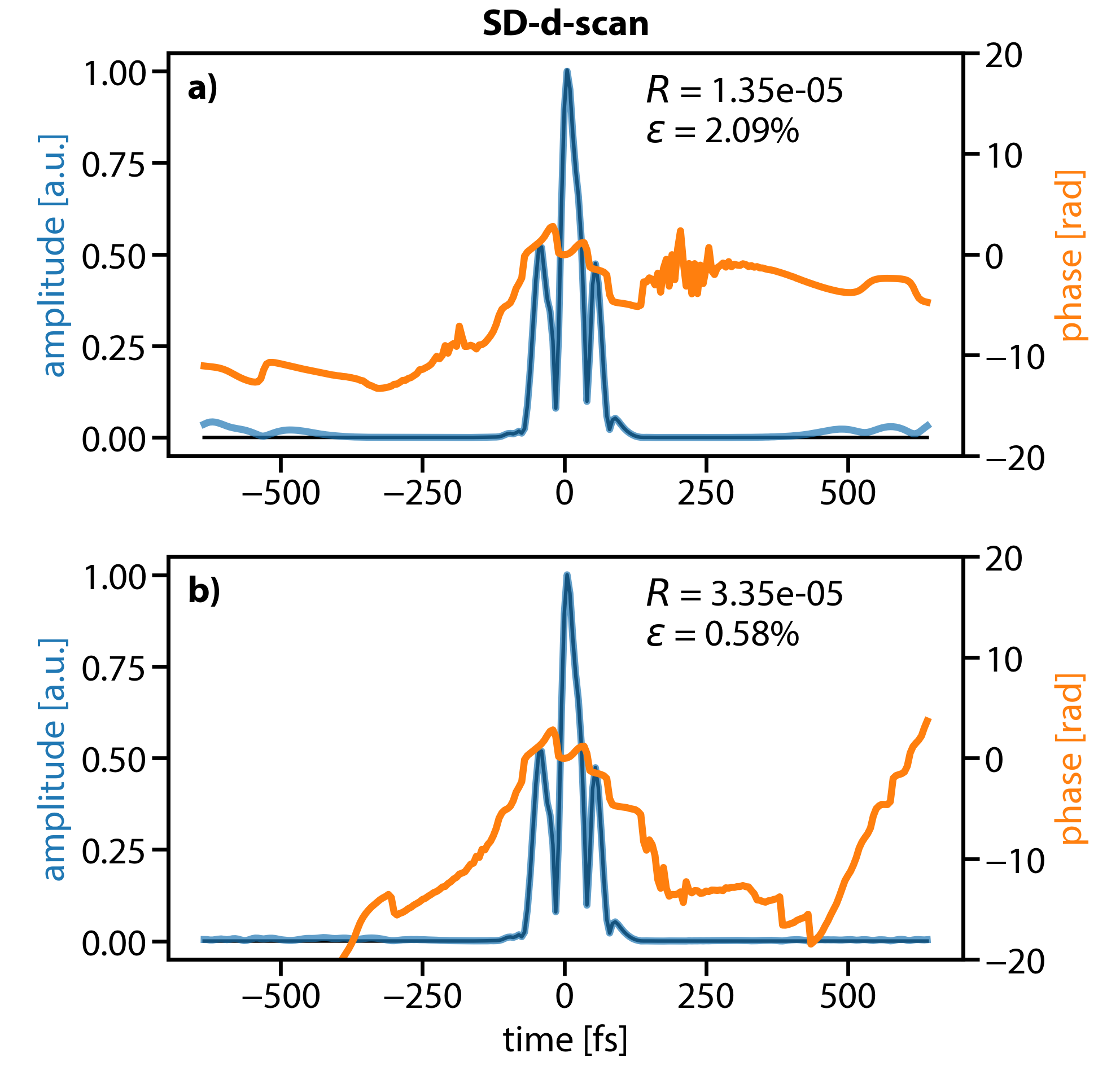}
    \caption{An example of the conditioning problems in SD-d-scan. Even though the pulse in a) has a lower trace error~$R$ than the one in b), it contains spurious satellite pulses and has a much higher retrieval error~$\epsilon$. The black line indicates the amplitude of the original test pulse.}
    \label{fig:illconditioned}
\end{figure}
As described in the main paper we performed a second retrieval simulation in which COPRA was initialized with a random guess (random amplitude and phase, uniformly distributed on $[0, 1]$ and $[0, 2\pi]$). Additionally, for d-scan and MIIPS the iteration count had to be increased from 300 to 1000 to obtain comparable retrieval errors. The retrieval ratios and the median retrieval error are shown in Fig.~\ref{fig:random_retrieval_probabilities}. We see that for many PNPS schemes COPRA performs almost equally well as with the Gaussian initial guess. Specifically, for the non-collinear methods~(FROG and TDP) there is no difference at all and for iFROG we only see minor differences in retrieval ratio and retrieval accuracy. This demonstrates that COPRA often works without any knowledge about the measured pulse.

For the d-scan and MIIPS we observe a different behavior. First of all, the retrieval ratio is lower, especially for MIIPS. The reason are local minima of the pulse retrieval problem to which the algorithm converges. Particularly many exist for MIIPS as the applied phase patterns are periodic, admitting local solutions that are spectrally shifted with respect to the global solution. Consequently, they do not appear when starting with a Gaussian pulse that is localized in time and frequency as they do not overlap much with this initial guess.

Secondly, the retrieval accuracy for MIIPS and d-scan is lower than when starting from Gaussian initial guesses. Strikingly, this can be seen for SD-d-scan and SD-MIIPS where even in the noiseless case a high retrieval error remains. The reason is that the retrieval problem is ill-conditioned. Fields at large delays compared to the main pulse, i.e., satellite pulses, contribute only weakly to the PNPS trace for these schemes~(see Fig.~\ref{fig:illconditioned}). The solutions obtained in this case have a small oscillating artifact in the frequency domain. They do not constitute non-trivial ambiguities in a strict sense as for noiseless measurements the correct solution can still be found when further reducing the trace error. However, COPRA stagnates when converging towards it. Curiously, this is not the case if the temporal grid is chosen too small and unphysical wrap-around happens in the calculation of the PNPS trace~(see Sec.~\ref{sec:test_pulses}).

One way to avoid this convergence behavior is to regularize the problem by, e.g., incorporating knowledge about the pulse. This was done implicitly in the main paper by starting from a Gaussian pulse in the time domain which does not contain the problematic contributions at large times. In any other case we found it is sufficient to simply set the first and last $10\%$ of the pulse field to zero after every iteration of COPRA. When testing for non-trivial ambiguities we excluded these artifacts by calculating a modified retrieval error $\epsilon'$ that takes only the central part of the pulse field $\vE$ into account.

\section{Large time-bandwidth products}\label{sec:high_tbp}
\begin{figure*}[tb]
    \centering
    \includegraphics{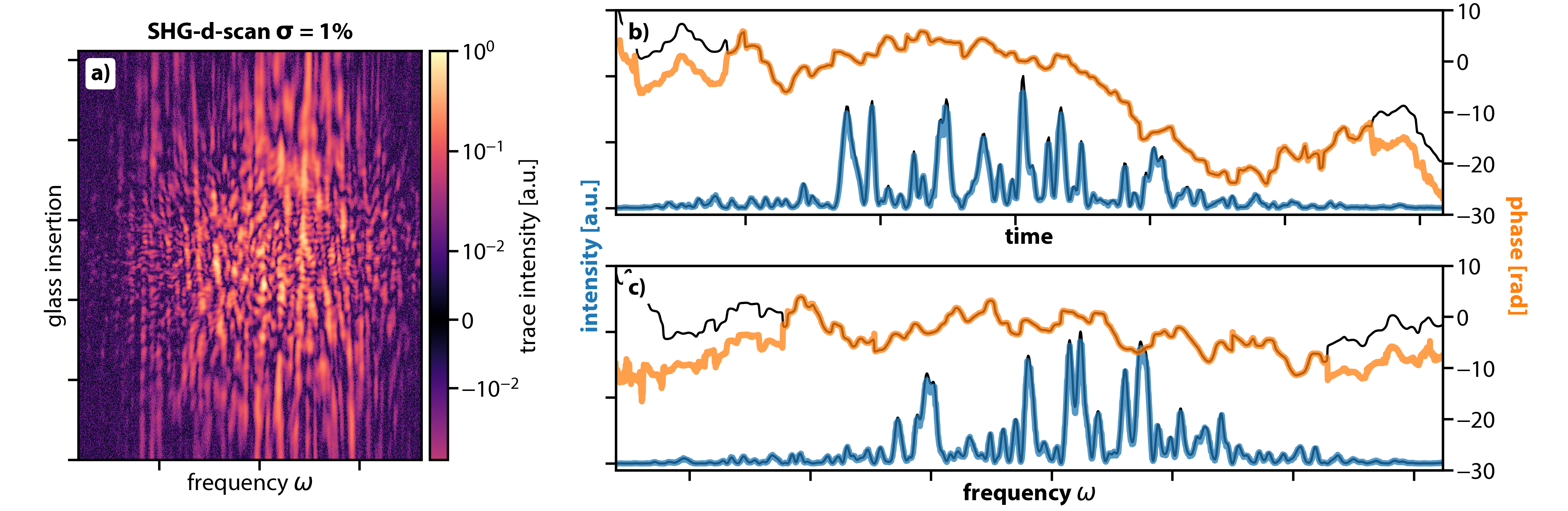}
    \caption{Example for COPRA retrieval of a very complex pulse with TBP 25 from a large synthetic SHG-d-scan measurement ($N=16384$, $M=512$) with a moderate level of Gaussian noise ($\sigma=1\%$). a) shows the noisy d-scan trace. b) and c) show the test pulse (black) and the retrieved solution (blue: intensity, orange: phase) in the time and frequency domain. The trace error of the solution is $R = \num{9.82e-3} = R_0 + \num{1.03e-7}$, the retrieval error is $\varepsilon=3.7\%$.}
    \label{fig:large_tbp}
\end{figure*}
Although not explicitly shown in the main paper, COPRA can, in principle, deal with measurements of pulses of arbitrary TBPs. In fact, in non-comprehensive numerical tests we found no significant dependence of convergence speed and retrieval probability on the TBP. One example for retrieval of a very complex pulse with TBP 25~($N=16384$) from a synthetic d-scan trace ($M=512$) is shown in Fig.~\ref{fig:large_tbp}. The retrieval with 150~iterations required roughly 5~min on an average notebook. Retrieval from measurements this large would probably be practically infeasible with state-of-the-art d-scan retrieval algorithms.

{\small
\bibliography{papers}}